\documentclass[10pt,final,journal,twoside]{IEEEtran}

\usepackage{bm}

\usepackage{cite}

\ifCLASSINFOpdf
  \usepackage[pdftex]{graphicx}
\else
  \usepackage[dvips]{graphicx}
\fi
\usepackage[cmex10]{amsmath}
\usepackage{algorithmic}

\usepackage{array}

\ifCLASSOPTIONcompsoc
  \usepackage[caption=false,font=normalsize,labelfont=sf,textfont=sf]{subfig}
\else
  \usepackage[caption=false,font=footnotesize]{subfig}
\fi
 \usepackage{dblfloatfix}
\usepackage{url}

\usepackage{lineno}

\usepackage{subeqnarray}
\usepackage{cases}
\usepackage{changebar}
\usepackage{amsfonts}
\usepackage{color}

\newtheorem{theorem}{Theorem}[section]

\newtheorem{definition}{Definition}[section]

\graphicspath{{fig/}}

\begin{document}


%
\title{Optimal ALOHA-like Random Access with Heterogeneous QoS Guarantees for Multi-Packet Reception Aided Visible Light Communications}

%
%

\author{
{Linlin~Zhao, Xuefen~Chi and Shaoshi~Yang, \IEEEmembership{Member,~IEEE} 
}%
\thanks{Manuscript received Mar 23, 2016; revised September 3, 2016.}
\thanks{L. Zhao and X. Chi are with the Department of Communication Engineering, Jilin University, Changchun China (e-mail: zhaoll13@mails.jlu.edu.cn; chixf@jlu.edu.cn).} 
\thanks{S. Yang is with the School of Electronics and Computer Science, University of Southampton, Southampton, SO17 1BJ, UK (e-mail: sy7g09@ecs.soton.ac.uk).} 
}

%
%

\markboth{Accepted to appear on IEEE Transactions on Wireless Communications, Sept. 2016}%
 {Zhao \MakeLowercase{\textit{et al.}}: Optimal ALOHA-like Random Access with Heterogeneous QoS Guarantees}
%



\maketitle
\begin{abstract}
There is a paucity of random access protocols designed for alleviating collisions in visible light communication (VLC) systems, where carrier sensing is hard to be achieved due to the directionality of light. To resolve the problem of collisions, we adopt the successive interference cancellation (SIC) algorithm to enable the coordinator to simultaneously communicate with multiple devices, which is referred to as the multi-packet reception (MPR) capability. However, the MPR capability could be fully utilized only when random access algorithms are properly designed. Considering the characteristics of the SIC aided random access VLC system, we propose a novel effective capacity (EC)-based ALOHA-like distributed random access algorithm for MPR-aided uplink VLC systems having heterogeneous quality-of-service (QoS) guarantees. Firstly, we model the VLC network as a conflict graph and derive the EC for each device. Then, we formulate the VLC QoS-guaranteed random access problem as a saturation throughput maximization problem subject to multiple statistical QoS constraints. Finally, the resultant non-concave optimization problem (OP) is solved by a memetic search algorithm relying on  invasive weed optimization and differential evolution (IWO-DE). We demonstrate that our derived EC expression matches the Monte Carlo simulation results accurately, and the performance of our proposed algorithms is competitive.
\end{abstract}


%
\IEEEpeerreviewmaketitle

\section{Introduction} 
The fifth generation (5G) wireless communication system focuses on achieving higher performance in a variety of technical aspects, such as area spectral efficiency,  delay, scalability, as well as reliability \cite{VLC-5G-2014}. 
However, radio frequency (RF)-based wireless communication technologies have arrived at a bottleneck to satisfy these requirements. 
Visible light communication (VLC) is recognized as an ideal complement to RF-based technologies in future 5G networks \cite{VLC-home} thanks to its advantages, such as high data rates, the avoidance of interference with RF systems, and low power consumption. 
Additionally, the 5G network is expected to carry machine-to-machine (M2M) traffic emerging from the Internet of Things (IoT). M2M communications may entail substantial uplink traffic having different delay-bounded quality-of-service (QoS) requirements \cite{5G-M2M-survey}.

At the time of writing, most existing studies of VLC focus on general-purpose indoor wireless communications, where it remains uncommon to use visible light for uplink transmissions. Nevertheless, with the development of M2M communications, there will be more and more scenarios where VLC becomes a competitive solution to the uplink transmission\cite{VLC-uplink-TDD, VLC-uplink-Arch, VLC-standard-15.7,VLC-Analysis-802.15.7, VLC-uplink-csmacd}, such as logistics centres, warehouses and indoor surveillance systems. In particular, there exist relevant application scenarios where the use of RF transmissions may be restricted, such as hospitals, chemical plants and airplanes. 
Therefore, VLC constitutes a promising alternative to RF based technologies for M2M communication in RF restricted areas\cite{VLC-M2M}, where the light-emitting diode (LED) indicator of machines/sensors can be used as optical transmitters for communicating with the coordinator.

As far as the existing VLC-based uplink communication schemes are concerned, the authors of \cite{VLC-uplink-TDD} experimentally demonstrated a time-division-duplex (TDD) VLC system for both downlink and uplink transmissions, and the authors of \cite{VLC-uplink-Arch} proposed a network architecture for a high-speed bi-directional VLC local area network (LAN) based on a star topology. An access protocol was also proposed in \cite{VLC-uplink-Arch} for the VLC network, where each indoor user transmits data with a RGB LED to the LED lamp in their individual pre-assigned time slot on the uplink\cite{VLC-uplink-Arch}. 
Furthermore, in 2011 the IEEE 802.15.7 standard \cite{VLC-standard-15.7} was presented, which defines the physical (PHY) layer and medium access control (MAC) layer for short-range wireless optical communications, including the scenario where all the links use visible light. With the aid of the Markov chain theory, the authors of \cite{VLC-Analysis-802.15.7} analysed the performance of the MAC protocol that relies on the carrier-sensing multiple-access with collision avoidance (CSMA/CA) mechanism and is used in the IEEE 802.15.7 protocol.  
In \cite{VLC-uplink-csmacd}, Wang \textit{et al.} proposed a MAC protocol relying on the CSMA with collision detection and hidden avoidance (CSMA/CD-HA) for bi-directional VLC wireless personal area networks, where each node solely uses a single LED to transmit and receive data. 
\par
Thanks to the simplicity and distributed feature, random access is suggested as one of multiple-access mechanisms by the IEEE 802.15.7 standard and other researchers \cite{VLC-standard-15.7,VLC-Analysis-802.15.7, VLC-uplink-csmacd}. 
However, it is challenging to address the issue of the competition among devices\footnote{In the IEEE 802.15.7 standard, devices refer to user terminals.} in VLC random access systems. 
Due to the directionality of light sources, the performance of VLC systems mainly depends on line-of-sight (LOS) transmissions \cite{VLC-LOS-channel}. As a result, the benefits of the random access based on carrier sensing are degraded. 
Some researchers have turned to the request-to-send/clear-to-send (RTS/CTS) mechanism and its modified versions for VLC systems \cite{VLC-LAST2015}.
However, RTS/CTS operates at the cost of frequent signalling. 
\par
There exists another alternative random access solution that is based on multi-packet reception (MPR). MPR is an approach embracing interference to obtain high throughput gain \cite{MPR-LangTong2013}. 
The receiver with MPR capability simultaneously decodes multiple signals transmitted from different devices. 
MPR can be implemented with the PHY layer techniques, such as successive interference cancellation (SIC), directional antennas and multiple-input-multiple-output (MIMO) techniques \cite{Shaoshi2015, LinlinZhao2015}. It can also be implemented with the ZigZag decoding \cite{MPR-zigzag} in the MAC layer. 
A few researchers have been exploring the redesign of MAC protocols for MPR-aided RF networks \cite{MPR-MAC-backoff2009,MPR-multi-round,MPR-Access-YunH2014,MIMO-CSMA2014}. 
The authors of \cite{MPR-MAC-backoff2009} stated that the optimal backoff factor of the CSMA/CA algorithm increases with the MPR capability. 
In \cite{MPR-multi-round}, a stop theory based RTS/CTS scheme was proposed to reduce the overhead of the RTS contention in MPR-aided systems. 
The authors of \cite{MPR-Access-YunH2014} proposed a flexible CSMA algorithm that adjusted the transmission probability according to the estimated number of active users for the non-saturation MPR-aided RF system. 
Relying on the zero-forcing based SIC (ZF-SIC) algorithm, the authors of \cite{MIMO-CSMA2014} proposed an analytical model to characterize the saturation throughput and mean access delay of CSMA/CA-aided RF networks. 
Nevertheless, none of the above works is dedicated to VLC systems, or to the scenario where the system is subject to delay QoS constraints. 
\par
There are indeed few contributions to the exploitation of MAC protocols in MPR-aided VLC systems. 
In \cite{MPR-fengfeng}, MPR was first explored for mitigating collisions in VLC random access systems. 
The authors analysed the impact of the MPR on the throughput performance of the VLC system. 
However, they did not consider the realistic effect of the PHY layer signal processing technology and did not redesign the MAC algorithm to fully utilize the MPR capability in the VLC system. 
As we mentioned before, due to the directionality of light sources, VLC systems are dramatically different from RF systems. As a result, it is critical to design a novel random access mechanism, especially the advanced random access mechanism with heterogeneous QoS guarantees, for MPR-aided VLC systems. 
\par
Based on the effective bandwidth (EB) concept \cite{EA1992}, Wu \textit{et al.} proposed the effective capacity (EC) theory \cite{EC2003}, where the concept of the statistical delay QoS, instead of a deterministic delay, is introduced to guarantee a specific delay bound violation probability. 
Focusing on the stochastic nature of RF fading channels, many researchers derived ECs of RF-based cellular systems and heterogeneous systems. Relying on the derived ECs, they developed their QoS-guaranteed resource allocation algorithms \cite{IWO-EC, EC-LTE2015, Hanzo2015, EC-performance2015}. 
Some of the above works focused on the downlink systems, while others focused on the uplink.
The works considering the uplink are all based on deterministic scheduling rather than random access. 
Due to the multi-fold stochastic nature, the random access system requires a new analysis framework of EC, which is the foundation for developing QoS-guaranteed random access algorithms. 
\par
In this paper, we focus on the uplink of a VLC system consisting of a coordinator and a number of devices. This is an important application scenario, since there will be a large amount of M2M traffic in the uplink of 5G systems. Similar to \cite{LinlinZhao2015} and \cite{MIMO-CSMA2014}, we employ the MIMO-SIC algorithm as the MPR technique to avoid collisions. 
For the VLC random access system with MIMO-SIC, the set of transmitting devices in each time slot is random, which means the group of the concurrently transmitting devices and their decoding order are both random. 
This characteristic is in stark contrast to that of conventional deterministic scheduling based cellular system and hence imposes significant challenges on redesigning a new random access algorithm. 
Inspired by the concept of EC, we propose a novel slotted ALOHA-like random access algorithm with heterogeneous delay QoS guarantees for the MPR-aided VLC system considered. 
The purpose of our algorithm is to make sure that each device accesses  the shared channel with its optimal probability according to its individual QoS constraint in a \textit{distributed} mode. 
Our ALOHA-like random access approach conceived for the scenario of heterogeneous QoS requirements can also be applied to the scenario of unified QoS requirements, which represents a special case of the former one. 
Our ALOHA-like random access algorithm is also suitable for networks using infrared (IR) for uplink transmissions, because of the similarity between the features of VLC channels and those of IR channels. Explicitly, 
to the best of our knowledge, the problem of designing a distributed random access based MAC protocol that takes into account the realistic effect of the adopted PHY layer signal processing technique and supports heterogeneous delay QoS requirements has not been investigated for MPR-aided VLC systems. Our main contributions are summarized as follows. 
\begin{itemize}
\item Considering the random nature of the access procedure of conflicting links, we model the ALOHA-like random access network as a conflict graph and propose the concept of \textit{feasible access state} for the MPR-aided system. 
Based on the conflict graph, we address the challenges of formulating the EC of the individual device taking into account the SIC algorithm, the random features of the system (randomly blocked channel and random access mechanism) as well as their interaction.  
\item Based on the EC derived, we propose a novel ALOHA-like random access algorithm for the MPR-aided VLC system under heterogeneous QoS guarantees. We derive the algorithm by solving the saturation throughput maximization problem subject to heterogeneous QoS constraints. 
\item For high-dimensional non-convex or non-concave constrained optimization problems (COPs), typically the optimum solution is difficult to obtain with the brute-force search due to multiple mathematical intractabilities, such as huge search space together with small feasible region, and the fact that the optimal solutions sometimes lie on constraint boundaries. 
In this paper, we address the non-convex/non-concave COP formulated using a novel memetic search algorithm that amalgamates the modified invasive weed optimization (IWO) algorithm relying on \textit{Pareto dominance} with the differential evolution (DE) algorithm, which helps us tackle the mathematical intractability of the COP considered.
\end{itemize}
\par
The rest of this paper is organized as follows. 
In Section \ref{s2}, the system model is described.  
In Section \ref{sec-HQoS}, the EC of individual device is derived and our  ALOHA-like random access algorithm with heterogeneous QoS guarantees is proposed. In Section \ref{sec-IWODE}, the IWO-DE algorithm is employed to solve the COP formulated. Our simulation results are provided in Section \ref{sec-results}. Finally, in Section \ref{sec-conclusion} the conclusions are offered. 
\par
\textit{Notation}: Bold letters denote vectors or matrices. 
$\textbf{1}_n$ denotes an $n \times 1$ vector whose elements are all $1$, and ${\textbf{1}}_{n \times n}$ denotes an $n \times n$ identity matrix. 
$(\bullet)^{H}$ denotes the Hermitian transpose, and $(\bullet)^{T}$ denotes the transpose. 
$\mathbb{E}(\bullet)$ denotes the expectation. 
$\{0,1\}^{N}$ denotes the set of all $N$-dimensional binary vectors. 
${\left\| \bullet \right\|}$ denotes the Euclidean vector norm. 
For the given $\sigma$, $N(0,\sigma^2)$ denotes the normal distribution with the mean of $0$ and the variance of $\sigma^2$. 
For the given vector $\textbf{x}$ and $\textbf{y}$, $\textbf{x} \prec \textbf{y}$ denotes that $\textbf{x}$ \textit{Pareto-dominates} $\textbf{y}$ for the given multi-objective optimization problem (MOP). 

\begin{figure}[tbp]
\centering
\includegraphics[width=3.5in]{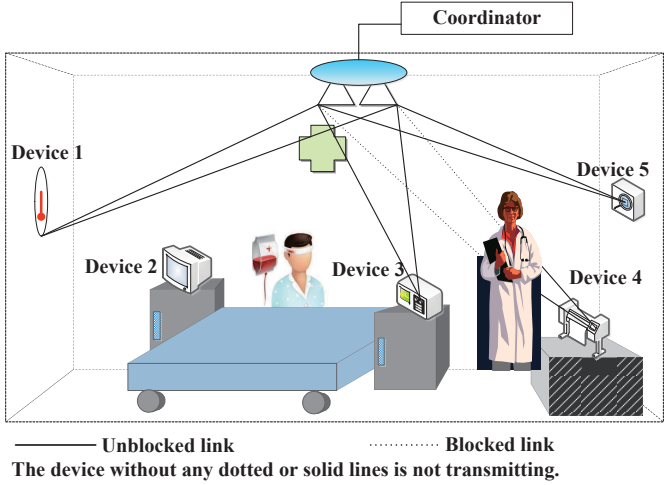}
\caption{The system model.}
\label{fig-system}
\end{figure}

\section{System Model}
\label{s2}
In this paper, we consider the uplink of an indoor MPR-aided VLC system with the star topology for M2M communications. The system model is shown in Fig. \ref{fig-system}. 
The system consists of one coordinator and $N$ devices. The coordinator is equipped with $M$ optical receivers (i.e., photo-detectors (PDs)), and each device is equipped with one optical transmitter (i.e., LED). The coordinator uses the MIMO-SIC algorithm to decode up to $M$ data streams simultaneously, i.e., the coordinator has the MPR capability of $M$. 
We assume that the coordinator knows the LOS channel gain information of devices. 
The time is divided into slots. 
At the beginning of each time slot, the device $j$ \textit{distributedly} transmits its data packets to the coordinator with an access probability of $p_j$, $j = 1, \cdots, N $, and its transmission occupies one time slot. 
We call this random access mechanism as the ALOHA-like random access mechanism. 
Moreover, packets are assumed constant-length. 
When the number of devices is not greater than the MPR capability of the coordinator, i.e., $N \leq M$, the coordinator is capable of decoding all the data streams of the devices, which is regarded as a situation where no collision happens.
Hence we mainly focus on the overloaded scenario that $N > M$.
\par
At time slot $t$, $n$ ($n \leq N$) unblocked devices are assumed to  transmit their signals which are organized into an $n \times 1$ non-negative real vector ${\bf{x}} = [x_1(t),\cdots, x_n(t)]^T$, and 
$\mathbb{E}\{ {\bf{x}} {\bf{x}}^T\}=\bm{1}_{n \times n}$. 
The received signal currents after the optical-to-electrical
conversions are collected into an $M \times 1$ vector $\textbf{r} = [r_1(t), \cdots, r_M(t)]^T$, which is given by
\begin{equation}
\label{eq-received-signal}
\textbf{r} = \xi P_\text{t} \bf{H}\bf{x}+\bf{v},
\end{equation} 
where $\xi$ is the detector responsivity, 
$P_\text{t}$ denotes the transmitting power, 
and $\bf{H}$ denotes the $M \times n$ indoor VLC channel matrix with element $h_{ij}$ representing the channel gain from the LED of device $j$ to the $i$-th PD of the coordinator, $i=1,\cdots,M$. Additionally, $\bf{v}$ denotes the additive white Gaussian noise (AWGN) with zero mean and variance $\sigma_\text{VLC}^2$, i.e., $\mathbb{E}\{ {\bf{v}} {\bf{v}}^H\}=\sigma_\text{VLC}^2 \bm{1}_{M \times M}$.
\par
In VLC systems, the reflected light signals are very small compared with the LOS light signals \cite{VLC-LOS-channel}. Therefore, we mainly consider the LOS transmission for convenience of analysis. 
The LOS channel gain $h_{ij}$ is given as \cite{VLC-LOS-channel}:
\begin{equation}
\label{eq-LOS-h}
h_{ij} = \left\{ {\begin{array}{*{20}{c}}
\displaystyle
{\frac{{(\rho + 1)A_0{G_\text{s}}}}{{2\pi {d_{ij}^2}}}{\varsigma({\psi ^{\text{in}}_{ij}})}{{\cos }^\rho}({\psi ^{\text{ir}}_{ij}}){\cos}({\psi ^\text{in}_{ij}}),{\psi ^\text{in}_{ij}} \le {\Psi _\text{C}}},\\
\displaystyle
{0,\quad \quad \quad \quad \quad \quad \quad \quad \quad \quad \quad \quad \quad \quad \quad  {\psi ^\text{in}_{ij}} > {\Psi _\text{C}}},
\end{array}} \right.
\end{equation}
where $\rho$ is the order of Lambertian emission, which is given by the semi-angle $\phi_{1/2}$ at half illumination of a LED as $\displaystyle \rho = \frac{{\ln 2}}{{\ln (\cos{\phi _{1/2}})}}$. 
$\Psi_\text{C}$ denotes the width of field of view (FOV) at a receiver.
$A_0$ is the physical area of the detector in a PD, and $G_\text{s}$ is the gain of an optical filter. 
$d_{ij}$ is the distance between the LED of the device $j$ and the $i$-th PD of the coordinator.
${\psi^\text{ir}_{ij}}$ is the angle of irradiation from the LED of the device $j$ to the $i$-th PD of the coordinator, $\psi^\text{in}_{ij}$ is the angle of incidence from the LED of device $j$ to the $i$-th PD of the coordinator, and ${\varsigma({\psi ^\text{in}_{ij}})}$ is the gain of an optical concentrator, 
\begin{equation}
{\varsigma({\psi ^\text{in}_{ij}})} = \left\{ {\begin{array}{*{20}{c}}
\displaystyle
{\frac{n_0^2}{{\sin }^2({\Psi _\text{C}})},\;{\psi ^\text{in}_{ij}} \le {\Psi _\text{C}}},\\
\displaystyle
{0,\quad \quad \quad \;\; {\psi ^\text{in}_{ij}} > {\Psi _\text{C}}},
\end{array}} \right. 
\end{equation}
where $n_0$ denotes the refractive index.

\par
Due to moving obstructions, the LOS propagation of the VLC system might be blocked. 
In this paper, we assume that the device $j$ is blocked if all the LOS links of the device $j$ are blocked.
We define a binary random variable (r.v.) $\kappa _j$ to characterize the event if the device $j$ is blocked or not. $\kappa _j=1$ represents that the device $j$ is unblocked, otherwise we have $\kappa _j=0$. 
We assume that $\kappa _j$ obeys the Bernoulli distribution \cite{Hanzo2015}: 
\begin{equation}
\label{eq-block}
f(\kappa _j) = \left\{ {\begin{array}{*{20}{c}}
{\,\quad \;\beta_j,{\kappa_j}  = 1,}\\
{1 - {\beta_j},{\kappa_j}  = 0,}
\end{array}} \right.
\end{equation}
where $\beta_j$ denotes the probability of the event that the device $j$ is unblocked. 
Because of the randomness and unpredictability of the blocked situation, we assume that the device is not aware if its link is blocked. 
\par
In VLC system, the dominant noise contribution is assumed to be the shot noise and the thermal noise \cite{VLC-LOS-channel}, i.e., we have
\begin{equation}
\label{eq-noise}
\sigma_\text{VLC}^2 = \sigma_\text{shot}^2+\sigma_\text{thermal}^2.
\end{equation}
According to \cite{VLC-LOS-channel}, the shot noise variance is given by
\begin{equation}
\label{eq-noise-shot}
\sigma_\text{shot}^2 = 2q \xi P_\text{r} B+2 q I_\text{bg} I_2 B,
\end{equation}
where $P_\text{r}$ denotes the total received optical power at the coordinator; 
$\xi$ is the detector responsivity, which is the same as that of (\ref{eq-received-signal}); $q$ denotes the charge of an electron, i.e., $q = 1.602 \times 10^{-19}$ coulombs; $B$ denotes the noise bandwidth; $I_\text{bg}$ denotes the background current; and $I_2$ denotes the second Personick integrals. In this paper, we assume that a field-effect-transistor (FET) transimpedance receiver \cite{Smith1980Receiver} is used.
The thermal noise variance is given by
\begin{equation}
\label{eq-noise-thermal}
\sigma_\text{thermal}^2 = \frac{8 \pi \varrho T}{G_0}\delta A_0 I_{2} B^2 + \frac{16 \pi^2 \varrho T \Gamma}{g_\text{m}} \delta^2 A_0^2 I_{3} B^3,
\end{equation}
where $\varrho$ is the Boltzmann’s constant, $T$ is the absolute temperature, $G_0$ is the open-loop voltage gain, $\delta$ is the fixed capacitance of the PD per unit area, $\Gamma$ is the FET channel noise factor, $g_\text{m}$ is the FET transconductance, and finally, $I_3$ denotes the third Personick integrals.
Here, we opt for the parameter values used in \cite{VLC-LOS-channel}:$I_\text{bg}=5.1 \times 10^{-3}$ [A], $I_2= 0.562$, $T=295$ [K], $G_0=10$, $g_\text{m} = 30$ [mS], $\Gamma = 1.5$, $\delta=112$ [pF/$\text{cm}^2$], $I_3 = 0.0868$.
\par
We define the QoS exponent vector ${\boldsymbol{\theta}}=[\theta_1,\cdots,\theta_N]$ to characterize the heterogeneous delay QoS in the VLC system. 
The QoS exponent of the device $j$ is denoted as $\theta_j$ that characterizes the steady state delay violation probability of the device $j$ such that ${\Pr} \left\{ {D_j \geq D_j^{\max}} \right\} \approx {e^{ - \theta_j \mu_j D_j^{\max}}}$, where $D_j$ denotes the steady state\footnote{More explicitly, it means the queue or buffer is in its steady state.} delay [second] of the device $j$, $D_j^{\max}$ denotes the delay bound (maximum tolerable delay) and $\mu_j$ is the fixed rate [bits/s] jointly determined by the arrival process and the service process \cite{EC2003}. 
It is apparent that the QoS exponent $\theta_j$ plays an important role here. Larger $\theta_j$ corresponds to more stringent statistical delay QoS constraint, while smaller $\theta_j$ implies looser statistical delay QoS requirements.
In this paper, we assume that the device $j$ chooses the value of its QoS exponent $\theta_j$ from the range of $10^{-10}$ to $1$.
\par
EC may be interpreted as the maximum constant arrival rate that can be supported by the service process of the system subject to the delay QoS constraint specified by the QoS exponent. 
For time-uncorrelated service processes, the EC of the device $j$ is given by \cite{IWO-EC}:
\begin{equation}
\label{eq-EC}
\mathsf{EC}_j(\theta_j) =  - \frac{1}{\theta_j }\ln \mathbb{E}\left[ {e^{-\theta_j s_j}} \right],
\end{equation}
where $\mathbb{E}[\bullet]$ denotes the expectation operator and $s_j$ denotes the service rate\footnote{The service rate represents the data [bits] communicated over the given time period.} of device $j$.
\section{The Novel ALOHA-like Random Access Algorithm with Heterogeneous QoS Guarantees}
\label{sec-HQoS}
Based on the EC theory, we propose a novel ALOHA-like random access algorithm for the MPR-aided VLC system subject to heterogeneous QoS constraints. 
For the ALOHA-like random access aided VLC system where the coordinator uses the MIMO-SIC algorithm, it is challenging to model the EC of each device due to the mutual dependence between the MIMO-SIC algorithm and the ALOHA-like random access mechanism. 
\subsection{The Effective Capacity of Individual Device}
Let $R_j(t)$ denote the instantaneously achievable transmission rate of the device $j$ at the time slot $t$, $j = 1,\cdots,N$. The fundamental part of modelling EC of the device $j$ is to find the probability distribution of $R_j(t)$. 
$R_j(t)$ and its probability distribution are influenced by the MIMO-SIC algorithm and the ALOHA-like random access algorithm that is characterised by the access probability vector $\textbf{p}$. 
To emphasize the effect of $\textbf{p}$ on EC, we define $\mathsf{EC}_j(\textbf{p};\theta_j)$ as the EC of device $j$ in the remainder part. 
It is extremely difficult to quantify the impact of the MIMO-SIC algorithm on $R_j(t)$ because of the randomness of the interference relations existing among the links. 
\par
For solving the above EC derivation problem, we model the ALOHA-like random access aided VLC network as a conflict graph $\cal G=(\cal{V},\cal{E})$, 
where $\cal V$ denotes the set of devices, $\cal E$ denotes the set of interference relations among links. There is an edge between two vertices in $\cal G$ if their links interfere with each other. 
The vector $\bm{\sigma}^i \in {\lbrace{0,1}\rbrace} ^N $ is defined as the state $i$ of $\cal G$, $i = 1,\cdots,2^N$, and $\sigma _j^i$ denotes the $j$-th element in $\bm{\sigma}^i$, $j = 1,\cdots,N$.
$\sigma _j^i = 1$ represents that the device $j$ transmits  signals to the coordinator and it is non-blocked in the state $\bm{\sigma}^i$; otherwise, we have $\sigma_j^i=0$. 
For example, the current state of $\cal G$ corresponding to the system of Fig. \ref{fig-system} is $[1,0,1,0,1]$\footnote{In the random access VLC system considered, not all of the devices transmit data packets to the coordinator in each time slot. In the particular time slot shown in Fig. 1, the device $2$ does not transmit data packets, thus the second element of the vector characterizing the current state of $\cal G$ is $0$.}.  
Since the device $j$ does not know if it is blocked before its transmission, we obtain:
\begin{equation}
\label{eq-sigma-1}
{\Pr} \lbrace{\sigma _j^i = 1}\rbrace = p_j \beta_j,
\end{equation}
\begin{equation}
\label{eq-sigma-0}
{\Pr} \lbrace{\sigma _j^i = 0}\rbrace = 1 - p_j \beta_j,
\end{equation}
where $p_j$ denotes the access probability of the device $j$. 
Note that $\sigma^i_j=1$ does not mean that the transmission of the device $j$ is successful because collisions might happen. 
\par
Because we adopt the slotted ALOHA-like random access mechanism, the states of $\cal G$ in different time slots are independent and identically distributed. The probability of the occurrence of the state $\bm{\sigma}^i$, defined as $\pi ({\boldsymbol{\sigma} ^i})$, is given by
\begin{equation}
\label{eq-pr-sigma}
\pi ({\boldsymbol{\sigma} ^i})=\prod\limits_{j = 1}^N {\left[ {(1 - \sigma _j^i)(1 - {p_j}{\beta_j}) + \sigma _j^i{p_j}{\beta_j}} \right]}.
\end{equation}
Note that herein $\pi$ is the steady state distribution probability rather than the mathematical constant that represents the ratio of a circle's circumference to its diameter.
\begin{definition}[Feasible Access State]
If $\tau_i \leq M $, where $\tau_i=\sum\limits_{j=1}^N {\sigma^i_j}$, the state $\bm{\sigma}^i$ is defined as a \textit{feasible access state} in the MPR-aided network with star topology.
\end{definition}
\begin{definition}[Infeasible Access State]
If $\tau_i > M $, the state $\bm{\sigma}^i$ is defined as an \textit{infeasible access state} in the MPR-aided network with star topology.
\end{definition}
\par
$\cal I$ is defined as the set of the \textit{feasible access states} of $\cal G$, and ${\cal I} = {\lbrace { \boldsymbol{\sigma} ^i: \tau_i \leq M} \rbrace}$. 
\par
In this paper, we adopt the MIMO-SIC algorithm to achieve MPR for the coordinator\footnote{In random access systems, the group of concurrently transmitting devices is random, and the coordinator does not even know the number of these devices, which imposes challenges on the channel estimation. As a result, the implementation and performance analysis of the MIMO-SIC algorithm are affected. In fact,  the number of simultaneously transmitting devices can be estimated by using the threshold based rank estimation algorithm, and then existing MIMO channel estimation algorithms can be used.}. That is, after the signal of one device is decoded, its signal is stripped away from the aggregate received signal before the signal of the next device is decoded \cite{Shaoshi2015,FWC2005}. 
We assume that no device is decoded successfully in the \textit{infeasible access state}, 
while in the \textit{feasible access state} $\bm{\sigma}^i$ we assume the SIC procedure of the MIMO-SIC algorithm is ideal, i.e., there is no error  propagation in the decoding process. 
The performance of the MIMO-SIC algorithm is dependant on the detection ordering method, which is represented by a $\tau_i \times \tau_i$ matrix $\bf{F}$. If and only if the signal of the device $j$ is decoded at the $l$-th layer, we have $F_{jl}=1$. Otherwise, we have $F_{jl}=0$. 
After using the detection ordering, the rearranged channel matrix and the transmitted signal are denoted as $\bf{\tilde{H}}$ and $\tilde{\bf{x}}$,  
which are given by $\tilde{\bf{H}}=\bf{H} \bf{F}$ and $\tilde{\bf{x}}=\bf{F} \bf{x}$, respectively. 
The LOS channel gain vector of the device $j$ is defined as ${\bf{h}}_j$ that is the $j$-th column of $\bf{H}$. The LOS channel gain vector of the device decoded at the $k$-th layer is defined as $\tilde{\bf{h}}_k$ that is the $k$-th column of $\bf{\tilde{H}}$. 
Additionally, $x_j$ is the transmitted signal of the device $j$, and $\tilde{x}_k$ is the signal decoded at the $k$-th layer. 
After decoding and removing $l-1$ signals, the residual received signal vector is given by ${\bf{r}}_l = \sum\limits_{k = l}^{\tau_i} \xi P_\text{t} \tilde{\bf{h}}_k \tilde{x}_k + {\bf{v}}$. If $F_{jl}=1$, we have ${\tilde{\bf{h}}}_l={\bf{h}}_j$ and ${\tilde{\bf{x}}}_l={\bf{x}}_j$, then 
\begin{equation}
{\bf{r}}_l = \xi P_\text{t} {\bf{h}}_j x_j + \sum\limits_{k = l+1}^{\tau_i} \xi P_\text{t} {\tilde{\bf{h}}}_k \tilde{x}_k + {\bf{v}}.
\end{equation}
Let ${\bf{z}}_l$ denote the output of the MIMO detector (e.g. ZF or minimum mean-squared error (MMSE) filter) at the $l$-th layer. ${\bf{W}}^l$ is defined as the filter matrix used at the $l$-th layer, and ${\bf{W}}^l_j$ denotes the $j$-th row in ${\bf{W}}^l$ \cite{Shaoshi2015}. 
At the $l$-th layer, the channel gains of the residual devices are organized into the residual channel matrix ${\bf{\tilde{H}}}_l=[\tilde{\bf{h}}_l,\cdots,\tilde{\bf{h}}_{\tau_i}]$. 
The $j$-th element of ${\bf z}_l$ is given by   
\begin{equation}
\label{eq-output-equalizer}
{\bf{W}}^l_j {\bf{r}}_l = \xi P_\text{t} {\bf{W}}^l_j {\bf{h}}_j x_j + {\bf{W}}^l_j \left[ \sum\limits_{k = l+1}^{\tau_i} \xi P_\text{t} {\tilde{\bf{h}}}_k \tilde{x}_k + {\bf{v}} \right].
\end{equation}
Hence, the signal-to-interference-plus-noise ratio (SINR) of the device $j$ in the \textit{feasible access state} $\bm{\sigma}^i$ is: 
\begin{equation}
\label{eq-snr-device}
\displaystyle
\gamma^i_j = \frac{\xi^2 P_\text{t}^2 \mathbb{E} \left[ {\left\| {\bf{W}}^l_j {\bf{h}}_j x_j \right\|}^2 \right]}{\mathbb{E} \left[ {\left\| {\bf{W}}^l_j \left( \sum\limits_{k = l+1}^{\tau_i} \xi P_\text{t} {\tilde{\bf{h}}}_k \tilde{x}_k + {\bf{v}} \right) \right\|}^2 \right]}.
\end{equation} 
If the ZF filter \cite{Shaoshi2015,FWC2005} is used, ${\bf{W}}^l_j$ is the $j$-th row of the matrix ${\left({{\bf{\tilde{H}}}_l}^H {\bf{\tilde{H}}}_l \right) }^{-1} {\bf{\tilde{H}}}_l^H$,
and the SINR of the device $j$ in the \textit{feasible access state} $\bm{\sigma}^i$ is: 
\begin{equation}
\label{eq-sinr-zf}
\gamma^i_j = \frac{\xi^2 P_\text{t}^2 {\Vert {\bf{W}}^l_j {\bf{h}}_j \Vert}^2} {\sigma_\text{VLC}^2 {\left\| {\bf{W}}^l_j {\bf 1}_M \right\|}^2 }.
\end{equation}
If the MMSE filter \cite{Shaoshi2015,FWC2005} is used, 
\begin{equation}
\label{eq-mmse-filter}
{\bf{W}}^l_j = {{\bf{h}}}_j^H{\left( \sum\limits_{k = l+1}^{\tau_i} \xi^2 P_\text{t}^2 {\tilde{\bf{h}}}_k {\tilde{\bf{h}}}_k^H +{\sigma_\text{VLC}^2}{\bf{1}}_{M \times M} \right) }^{-1},
\end{equation}
and the SINR of the device $j$ in the \textit{feasible access state} $\bm{\sigma}^i$ is given by
\begin{equation}
\label{eq-sinr-mmse}
\gamma^i_j = \xi^2 P_\text{t}^2 {\bf{h}}_j^H{ \left( \sum\limits_{k = l+1}^{\tau_i}\xi^2 P_\text{t}^2 {\tilde{\bf{h}}}_k {\tilde{\bf{h}}}_k^H+{\sigma_\text{VLC}^2} {\bf{1}}_{M \times M} \right) }^{-1}{\bf{h}}_j.
\end{equation}
\par
Using Shannon's capacity formula, the upper bound on the instantaneously achievable transmission rate [bits/s] of the device $j$ in the \textit{feasible access state} ${\boldsymbol{\sigma} ^i}$ is given by
\begin{equation}
\label{eq-R-device}
R_j^i = B \times \log_2 \left( { 1+ \gamma^i_j} \right) \;\; \text{[bits/s]}.
\end{equation} 
\par
The instantaneously achievable transmission rate of the device $j$ only depends on the state of $\cal{G}$ (i.e., the set of the devices that are simultaneously transmitting signals) and their corresponding channel gains. The individual channel gains of these devices can be regarded as fixed provided that these devices are unblocked and their positions are unchanged. Because the states of $\cal{G}$ in different time slots are independent and identically distributed, the instantaneously achievable transmission rates of the device $j$ are independent and identically distributed across different time slots.
The probability of $\left\{ R_j(t) = R_j^i \right\}$ equals the probability of the occurrence of the \textit{feasible access state} $\bm{\sigma}^i$ that satisfies $\sigma_j^i=1$, and is given by
\begin{equation}
\label{eq-Pr-Rj-Rji}
{\Pr}\left\{ R_j(t) = R_j^i \right\} = \pi(\boldsymbol{\sigma} ^i), {\boldsymbol{\sigma} ^i} \in {\cal{I}}_j,
\end{equation}
where ${\cal{I}}_j$ is defined as the set of the \textit{feasible access states} satisfying that the device $j$ is transmitting signals to the coordinator and is unblocked, ${\cal{I}}_j = {\lbrace { \boldsymbol{\sigma} ^i: \tau_i \leq M, \sigma_j^i=1} \rbrace}$.
Furthermore, the probability of $\left\{ R_j(t) = 0 \right\}$ equals the probability of all events that the device $j$ transmits unsuccessfully, which is given by
\begin{equation}
\label{eq-Pr-Rj-0}
\begin{array}{l}
{\Pr}\left\{ R_j(t) =0 \right\} = { 1-\sum\limits_{{\boldsymbol{\sigma} ^i} \in {\cal{I}}_j} \pi(\boldsymbol{\sigma} ^i)} \\
\qquad\qquad\qquad\;\; = 1-\sum\limits_{{\boldsymbol{\sigma} ^i} \in \cal{I}} \sigma_j^i \pi(\boldsymbol{\sigma} ^i).
\end{array}
\end{equation} 
Accordingly, the EC of the device $j$ is formulated as (\ref{eq-ec-device-bits}).

\newcounter{MYtempeqncnt}
\begin{figure*}[t]
\normalsize
\hrulefill
\begin{equation}
\label{eq-ec-device-bits}
\begin{array}{l}
\displaystyle
\mathsf{EC}_j({\textbf{p}};\theta_j) = { - \frac{1}{\theta_j} \ln \left\{  \left[ \sum\limits_{{\boldsymbol{\sigma} ^i} \in {\cal{I}}_j} \pi(\boldsymbol{\sigma} ^i) e^{-\theta_j R_j^i} \right] + 1- \left[ \sum\limits_{{\boldsymbol{\sigma} ^i} \in \cal{I}} \sigma_j^i \pi(\boldsymbol{\sigma} ^i) \right] \right\} }\\
\vspace{1mm}
\displaystyle 
\quad \quad \quad \quad \;\; = - \frac{1}{\theta_j} \ln \left\{  \left[ \sum\limits_{{\boldsymbol{\sigma} ^i} \in \cal{I}} \sigma_j^i \pi(\boldsymbol{\sigma} ^i) e^{-\theta_j R_j^i} \right] + 1- \left[ \sum\limits_{{\boldsymbol{\sigma} ^i} \in \cal{I}} \sigma_j^i \pi(\boldsymbol{\sigma} ^i) \right] \right\} \\
\displaystyle 
\quad \quad \quad \quad \;\; =  - \frac{1}{\theta_j} \ln \left\{ 1- {\sum\limits_{{\boldsymbol{\sigma} ^i} \in \cal{I}}} \left[ \left( 1-e^{-\theta_j R_j^i} \right) \sigma_j^i {\prod\limits_{j = 1}^N {\left[ {(1 - \sigma _j^i)(1 - {p_j}{\beta_j}) + \sigma _j^i{p_j}{\beta_j}} \right]}} \right] \right\} \text{[bits/s]}
\end{array}
\end{equation}
\hrulefill
\vspace*{2pt}
\end{figure*}

\subsection{Problem Formulation}
Saturation throughput refers to the network throughput when devices always have data to transmit \cite{MIMO-CSMA2014}. According to (\ref{eq-Pr-Rj-Rji}), the saturation throughput is formulated as
\begin{equation}
\label{eq-saturation-throughput}
\begin{array}{l}
\eta({\textbf{p}}) = \sum\limits_{{\boldsymbol{\sigma} ^i} \in \cal{I}}  \sum\limits_{j=1}^{N} R_j^i \sigma_j^i \pi({\boldsymbol{\sigma} ^i}) \\
\qquad = \sum\limits_{{\boldsymbol{\sigma} ^i} \in \cal{I}} \sum\limits_{j=1}^{N} R_j^i \sigma_j^i {\prod\limits_{k = 1}^N {\left[ {(1 - \sigma _k^i)(1 - {p_k}{\beta_k}) + \sigma _k^i{p_k}{\beta_k}})\right] }} .
\end{array}
\end{equation}
\par
In this paper, we aim to find the optimum access probability vector ${\textbf{p}} = [p_1,\cdots,p_N]$ so that the saturation throughput is maximized for the VLC system subject to the heterogeneous delay QoS constraints. 
For a given QoS exponent vector $\boldsymbol{\theta} = \left[{\theta_1,\cdots,\theta_N}\right]$, the optimization problem (OP) is formulated as
\begin{equation}
\label{eq-HQoS-OP}
\begin{array}{l}
\displaystyle
{\mathop {\max }\limits_{\textbf{p}} \;\eta({\textbf{p}})  = \sum\limits_{{\boldsymbol{\sigma} ^i} \in \cal{I}} \sum\limits_{j=1}^{N} R_j^i \sigma_j^i \pi({\boldsymbol{\sigma} ^i})}\\
\displaystyle
{ \text{s.t.}\;\;\;\mathsf{EC}_j(\textbf{p};\theta_j) \ge \mathsf{EB}_j(\theta_j)},\\
\displaystyle
{\quad \quad 0 < {p_j} \le 1,\;j = 1, \cdots, N },
\end{array}
\end{equation}
where $\mathsf{EB}_j(\theta_j)$ denotes the EB \cite{EA1992} of the traffic of the device $j$, given by 
\[
\displaystyle \mathsf{EB}_j(\theta_j ) = \frac{1}{\theta_j } \mathop {\lim }\limits_{t \to \infty } \frac{1}{t} \ln \mathbb{E} \left[ {e^{\theta_j A_j(t)}} \right],
\] 
in which $A_j(t)$ denotes the accumulated arrival process of the device $j$ over the time interval $[0, t)$. 
$\mathsf{EC}_j({\textbf{p}};\theta_j) \ge \mathsf{EB}_j(\theta_j)$ represents that the QoS constraint of device $j$ has to be satisfied \cite{IWO-EC}. 
$0 < {p_j} \le 1$ represents that the access probability of device $j$ must range from $0$ to $1$. 
$\theta_j = 0$ denotes that the device $j$ has no QoS requirement.
\begin{theorem}
\label{theorem-non-convex}
The OP (\ref{eq-HQoS-OP}) is non-concave with respect to the access probability vector ${\textbf{p}}=\left[ {p_1,\cdots,p_{_N}} \right]$.
\end{theorem}
\begin{IEEEproof}
Please see Appendix \ref{app-theorem-non-convex}.
\end{IEEEproof}
\par
It is usually unlikely to transform a non-concave problem to a convex or concave problem without solution gap. As an evolutionary algorithm, IWO \cite{IWO-2006} provides an alternative to search for the optimum solution of  the unconstrained optimization problem (UOP), and a number of its variants have also been proposed for solving constrained optimization problems (COP). Unlike in the UOP, both the constraints and the objective function need to be considered to find the optimum solution of the COP. 
More recently, it is increasingly popular to solve COP using \textit{Pareto dominance} based multi-objective optimization (MOP) techniques.
In this paper, we transform the COP (\ref{eq-HQoS-OP}) into an MOP and adopt the emerging memetic algorithm called IWO-DE \cite{IWO-DE2013} to tackle the problem. 
For convenience, we rewrite (\ref{eq-HQoS-OP}) as follows
\begin{equation}
\label{eq-HQoS-OP-min}
\begin{array}{l}
\displaystyle
{\mathop {\min }\limits_{\textbf{p}} \;\frac{1}{\eta({\textbf{p}}) } = \frac{1}{\sum\limits_{{\boldsymbol{\sigma} ^i} \in \cal{I}} \sum\limits_{j=1}^{N} R_j^i \sigma_j^i \pi({\boldsymbol{\sigma} ^i})}}\\
\displaystyle
{\text{s.t.}\;\;\;\frac{{\mathsf{EB}_j({\theta _j})}}{{\mathsf{EC}_j({\textbf{p}};{\theta _j})}} \leq 1,}\\
{\quad \quad 0 < {p_j} \le 1,\; j = 1, \cdots, N }.
\end{array}
\end{equation}
Obviously, the OP (\ref{eq-HQoS-OP-min}) and the OP (\ref{eq-HQoS-OP}) share the same optimum solution and the same feasible solution set. 

\section{The IWO-DE algorithm}
\label{sec-IWODE}
The IWO-DE algorithm first considers IWO as a local refinement procedure to tackle the COP. 
The outputs of the IWO procedure are taken as the optimum candidates used as the inputs of the DE procedure, which is employed as the global search procedure to effectively explore the search space for finding the optimum access probability vector. 
A detailed discussion of IWO-DE is out of the scope of this paper. The interested reader is referred to \cite{IWO-DE2013}. 
Below we start with a brief introduction of the IWO algorithm. 
\subsection{IWO-The Local Refinement Procedure}
The classical IWO algorithm \cite{IWO-2006} cannot handle the COP directly, 
thus we transform the single-objective COP (\ref{eq-HQoS-OP-min}) into a bi-objective unconstrained OP (a special case of MOP, which is also called vector optimization problems \cite{book-convex}). 
We employ the modified IWO algorithm relying on \textit{Pareto optimization} technologies to cope with the bi-objective OP. 
Referring to \cite{IWO-DE2013}, the unconstrained bi-objective OP is given as the element-wise minimization problem of
\begin{equation}
\label{eq-bi-OP}
\mathop {\min }\limits_{\textbf{p}} \;\bm{\Lambda}({\textbf{p}}) = \left( {\frac{1}{\eta ({\textbf{p}})},\Omega({\textbf{p}})} \right),
\end{equation}
where the first objective $\displaystyle \frac{1}{\eta ({\textbf{p}})}$ is the original objective in the OP (\ref{eq-HQoS-OP-min}), and the second objective $\Omega({\textbf{p}})$ is the sum of all constraint violations in the OP (\ref{eq-HQoS-OP-min}). More specifically, $\Omega({\textbf{p}})$ is defined as follows:
\begin{equation}
\label{eq-Omega}
\Omega({\textbf{p}}) = \sum\limits_{j = 1}^N {\left[ {{\Omega}_j^{\text{QoS}}({\textbf{p}}) + {\Omega}_j^{p0}({\textbf{p}}) + {\Omega}_j^{p1}({\textbf{p}})} \right]},
\end{equation}
\begin{equation}
\label{eq-GQoS}
{\Omega}_j^{\text{QoS}}({\textbf{p}}) = \max \left( {0, \frac{{\mathsf{EB}_j({\theta _j})}}{{\mathsf{EC}_j({\textbf{p}};{\theta _j})}}-1 } \right),
\end{equation}
\begin{equation}
\label{eq-Gp0}
{\Omega}_j^{p0}({\textbf{p}}) = \max \left( {0, - {p_j}} \right),
\end{equation}
\begin{equation}
\label{eq-Gp1}
{\Omega}_j^{p1}({\textbf{p}}) = \max \left( {0,{p_j} - 1} \right).
\end{equation}
Obviously, we have $\Omega({\textbf{p}}) \ge 0$, and all the constraints in (\ref{eq-HQoS-OP-min}) are satisfied if and only if $\Omega({\textbf{p}}) = 0$. 
\par
The optimization variable $\textbf{p}$ of the problem (\ref{eq-bi-OP}), i.e. the access probability vector in this paper, is taken as the position of the weed in the search space.
The number of devices in the VLC system is $N$, hence $\textbf{p}$ is an $N$-dimensional variable. As a result, the search space of our IWO-DE algorithm is $N$-dimensional. Without causing confusion, in what follows both the position of the weed and the weed itself may represent the access probability vector. 
The candidate access probability vector that has smaller fitness value, as define in (\ref{eq-fitness}), is closer to the optimum access probability vector and thus more likely to reproduce and survive. 
For simultaneously achieving minimization of the two objectives in (\ref{eq-bi-OP}), some modifications are made in the \textit{reproduction} step and the \textit{competitive exclusion} step of the IWO algorithm. 
The key steps of the modified IWO algorithm are presented as follows. 
\par
Step 1) \textit{Initialization}: 
A number of initial candidate solutions defined as ${\cal P}= \lbrace {{\textbf{p}}}^i, i = 1,\cdots,W_0 \rbrace$ are randomly dispersed over the $N$-dimension space. $W_0$ denotes the number of the initial candidate solutions. ${\textbf{p}}^i$ is the $i$-th weed, i.e., the $i$-th candidate solution of the problem (\ref{eq-bi-OP}), in the population, and it is an $1 \times N$ vector with element $p_j^i$ representing the access probability of the device $j$. 
\par
Step 2) \textit{Reproduction}: 
To accommodate IWO for the MOP, we adopt the adaptive weighted sum fitness assignment mechanism \cite{IWO-DE2013} to determine the number of the offspring reproduced by each weed. 
For the weed $\textbf{p}^i$, the fitness value of ${{\textbf{p}}^i}$, denoted  as ${fit({{\textbf{p}}^i})}$, is defined as follows:
\begin{equation}
\label{eq-fitness}
fit({\textbf{p}}^i) = \sqrt {\omega \Delta_f{{({\textbf{p}}^i)}^2} + (1 - \omega ){\Delta_\Omega}{{({\textbf{p}}^i)}^2}}, 
\end{equation}
where
\begin{equation}
\label{eq-fit-weight}
\omega = \frac{\text{The\;size\;of\;feasible\;weeds}}{\text{The\;size\;of\;all\;weeds}},
\end{equation}
\begin{equation}
\label{eq-weight-object}
\Delta_f({\textbf{p}}^i) = \frac{{f({\textbf{p}}^i) - \mathop {\min }\limits_{j = 1, \cdots ,W}f({\textbf{p}}^j)}}{\mathop {\max }\limits_{j = 1, \cdots ,W}f({\textbf{p}}^j)-\mathop {\min }\limits_{j = 1, \cdots ,W}f({\textbf{p}}^j)},
\end{equation}
\begin{equation}
\label{eq-weight-constraints}
{\Delta_\Omega}({\textbf{p}}^i) = \frac{{\Omega({\textbf{p}}^i) - \mathop {\min }\limits_{j = 1, \cdots ,W}\Omega({\textbf{p}}^j)}}{{\mathop {\max }\limits_{j = 1, \cdots ,W}\Omega({\textbf{p}}^j) - \mathop {\min}\limits_{j = 1, \cdots ,W}\Omega({\textbf{p}}^j)}},
\end{equation}
in which we have $\displaystyle f({\textbf{p}}^i) = \frac{1}{\eta({\textbf{p}}^i)}$ and $W=W_0$ if $Z=0$. Otherwise, $W=W_{\max}$. 
$Z$ denotes the iteration index, and also denotes the generation index. 
$W_{\max}$ denotes the maximum number of survival weeds. 
\par
Hence, the number of offspring reproduced by the $i$-th weed ${{\textbf{p}}^i}$ is formulated as follows \cite{IWO-2006}:
\begin{equation}
\label{eq-seed-num}
{S_{i}} = {S_{\max} - \frac{(S_{\max} - S_{\min}){(fit({{\textbf{p}}}^i) - \mathop {\min }\limits_{j = 1, \cdots ,W}fit({\textbf{p}}^j))}}{{\mathop {\max }\limits_{j = 1, \cdots ,W}fit({\textbf{p}}^j) - \mathop {\min }\limits_{j = 1, \cdots ,W}fit({\textbf{p}}^j)}}} ,
\end{equation}
where 
$S_{\max}$ denotes the permissible maximum number of offspring, and $S_{\min}$ denotes the permissible minimum number of offspring. 
\par
Step 3) \textit{Spatial Dispersion}: 
The generated offspring of the $i$-th weed ${\textbf{p}}^i$ are randomly dispersed over the search space. 
The offspring obey the normal distribution with varying standard deviations. 
The mean of the normal distribution is zero, 
and its standard deviation gradually reduces from an initial value $\sigma_{\text{initial}}$ to a final value $\sigma_{\text{final}}$ in each generation. 
The standard deviation at each iteration is expressed as \cite{IWO-2006}:  
\begin{equation}
\label{eq-sd-iter}
{\sigma _{Z}} = {\left( {\frac{{Z_{\max } - Z}}{{Z_{\max }}}} \right)^{\phi}}\left( {{\sigma _\text{initial}} - {\sigma _\text{final}}} \right) + {\sigma _\text{final}},
\end{equation} 
where $Z_{\max}$ is the maximum number of iterations, $\sigma _{Z}$ is the standard deviation in the $Z$-th iteration and ${\phi}$ is the \textit{non-linear modulation index}\footnote{In the IWO algorithm, the \textit{non-linear modulation index} ${\phi}$ is a terminology that determines the variation degree of the offspring.} of the IWO algorithm \cite{IWO-2006}.
\par
Step 4) \textit{Competitive Exclusion}: 
In order to limit the number of the weeds that serve as inputs to the DE procedure, weeds and their offspring are ranked together to eliminate weeds that have poor (i.e. high herein) fitnesses values in this step.  
More specifically, to minimize the two objectives in (\ref{eq-bi-OP}) simultaneously, we adopt the non-dominated sorting \cite{IWO-DE2013} to decide which individual to survive into the next generation.  
The non-dominated sorting is based on \textit{Pareto dominance} \cite{IWO-DE2013}, 
which ensures that the candidate solutions obtained in the IWO procedure distribute surrounding the \textit{Pareto front} \cite{IWO-DE2013}. 
As a beneficial result, the convergence of the algorithm is accelerated.  
The definition of \textit{Pareto dominance} is given as follows \cite{IWO-DE2013}: 
\begin{definition}[Pareto Dominance]
For the given MOP with $m$ optimization variables ${\textbf{x}} = [{x_1}, \cdots ,{x_m}]$ and $n$ objective functions $g_i({\textbf{x}}), i \in \{ {1, \cdots, n} \}$, define
\begin{equation}
\label{eq-MOP}
\min \;{\textbf{y}} = \left( {{g_1}({\textbf{x}}), \cdots ,{g_n}({\textbf{x}})} \right).
\end{equation}
An optimization variable ${\textbf{x}}^a$ is said to \textit{Pareto-dominate} another vector ${\textbf{x}}^b$ (denoted by ${{\textbf{x}}^a}\prec{{\textbf{x}}^b}$) if and only if:
\[\forall i \in \{1,\cdots,n\}:g_i({\textbf{x}}^a) \le g_i({\textbf{x}}^b)\]
\begin{equation}
and \;\; \exists j \in \{1,\cdots,n\}:g_j({\textbf{x}}^a) < g_j({\textbf{x}}^b).
\end{equation}
\end{definition}
Accordingly, the exclusion mechanism of IWO is as follows:
\begin{itemize}
\item if the weed ${{\textbf{p}}}^i$ \textit{Pareto-dominates} the weed ${{\textbf{p}}}^j$, the weed ${{\textbf{p}}}^i$ wins;
\item otherwise, the weed with smaller \textit{sum constraint violation value} $\Omega(\textbf{p})$ wins.
\end{itemize}
\par
The top $W_{\max}$ weeds are taken as the optimum candidates used as the inputs of the DE procedure. 
Having carried out the \textit{refinement} procedures of IWO, the excellent weeds are picked out. Then, the DE algorithm is applied with respect to the selected weeds, which essentially explores the search space to obtain the optimum solution. 
\subsection{DE-The Global Search Model}
The \textit{variation} of the DE procedure is created to find the global optimum solution by means of three operations, i.e., \textit{mutation}, \textit{crossover} and \textit{selection} \cite{IWO-DE2013}.
\par
Step 1) \textit{Mutation Operation}:
In the mutation operation, a mutant vector ${\bm{\alpha}^i}$ is built  as follows \cite{IWO-DE2013}: 
\begin{equation}
\label{eq-DE-mutation}
{\bm{\alpha}^i} = {{\textbf{p}}^{\text{best}}} + F_0 \times \left( {{{\textbf{p}}^{{r_1}}} - {{\textbf{p}}^{{r_2}}}} \right),
\end{equation}
where ${\textbf{p}^{\text{best}}}$ denotes the best weed obtained in the IWO procedure, $r_1$ and $r_2$ denote randomly selected indices, 
$\textbf{p}^{r_1}$ denotes the $r_1$-th weed in the IWO procedure, and $\textbf{p}^{r_2}$ denotes the $r_2$-th weed in the IWO procedure.
$F_0$ is a scaling factor that quantifies the difference between $\textbf{p}^{r_1}$ and $\textbf{p}^{r_2}$. 
Each weed in the IWO procedure corresponds to a mutant vector in the DE procedure. 
$\bm{\alpha}^i$ is a $1 \times N$ vector, $i=1,\cdots,W_{\max}$.
\par
Step 2) \textit{Crossover Operation}:
With the mutant vector $\bm{\alpha}^i$ and the vector $\textbf{p}^i$, another trial vector $\textbf{u}^i$ is generated through the binomial crossover as follows \cite{IWO-DE2013}:
\begin{equation}
\label{eq-DE-crossover}
u_j^i = \left\{ {\begin{array}{*{20}{c}}
{\alpha_j^i,\text{if}\;\text{rand}_j \le {C_\text{r}},\text{or} \;j = {j_\text{rand}}}\\
{p_j^i,\text{otherwise} \quad \quad \quad \quad \quad \quad \quad}
\end{array}}, \right.
\end{equation}
where $i=1,\cdots,W_{\max}$, $j=1,\cdots,N$, $j_\text{rand}$ is a randomly selected integer ranging from $1$ to $N$, which ensures $\textbf{u}^i$ inherits at least one component from the mutant vector $\bm{\alpha}^i$. Additionally, $\text{rand}_j$ is a r.v. obeying the uniform distribution and $0 \leq \text{rand}_j \leq 1 $. $C_\text{r}$ is the crossover probability. $\alpha_j^i$, $p_j^i$ and $u_j^i$ denote the $j$-th element in $\bm{\alpha}^i$, $\textbf{p}^i$ and $\textbf{u}^i$, respectively.
\par
Step 3) \textit{Selection Operation}:
$\textbf{u}^{i}$ is chosen as the replacements of $\textbf{p}^{i}$ in the next generation only when ${\textbf{u}}^{i}$ leads to a better solution. Hence, the selection operation is described by:
\begin{equation}
\label{eq-DE-selection}
{{\textbf{p}}^{i}} = \left\{ {\begin{array}{*{20}{c}}
{{{\textbf{u}}^{i}},\text{if}\;{{\textbf{u}}^{i}} \prec {{\textbf{p}}^{i}}}\\
{{{\textbf{p}}^{i}},\text{otherwise},}
\end{array}} \right.
\end{equation} 
For the sake of clarity, the whole framework of IWO-DE is presented in Table \ref{table-IWO-DE}. 
The search space of our IWO-DE is $N$-dimensional. Hence, the computational complexity of the IWO-DE algorithm is ${\cal{O}}\left[ Z_{\max} W_{\max} S_{\max} ( N  + W_{\max} S_{\max} ) \right]$. A larger maximum number of weeds ($W_{\max}$) and a higher maximum number of offspring ($S_{\max}$) may improve the quality of the optimum access probability vector obtained, while imposing an increased computational complexity.
\begin{table}[!htbp]
\centering
\renewcommand{\arraystretch}{1.2}
\caption{The IWO-DE Algorithm}
\label{table-IWO-DE}
\vspace*{4pt}
\begin{tabular*}{8.6cm}{l}
\hline
\% Initialization\\
$Z=0$;\\
\% Create initial population of $W_0$ weeds ${\cal P}= \lbrace {\textbf{p}}^i, i=1,\cdots,W_0 \rbrace$ \\
\% that are randomly dispersed over the $N$-dimensional search space,\\
\%  and the range of the value in each dimension is between $0$ and $1$;\\
\hline
\vspace{2pt}
\textbf{while} $Z<{Z}_{max}$\\
\qquad \%\% (\textbf{Operate the IWO algorithm})\\
\qquad \% Evaluate the fitness value of each weed in ${\cal P}$ according to (\ref{eq-fitness});\\
\qquad \% Sort ${\textbf{p}}^i$ in ascending order according to their fitness values;\\
\qquad \% For each weed ${\textbf{p}}^i$, from (\ref{eq-seed-num}), calculate the size of its offspring $S_i$.\\
\qquad \% For each ${{\textbf{p}}^i}$, create its offspring set ${{\cal P}_{s_i}}=\lbrace {{\textbf{p}}_{s_i}^k, k = 1,\cdots,S_i} \rbrace$, \\
\qquad \textbf{for} $k = 1:S_i$\\
\qquad\qquad ${\textbf{p}}_{s_i}^k = {\textbf{p}}^i + {{\boldsymbol{\varphi}} ^k}$,\\
\qquad\qquad \% all elements in ${\boldsymbol{\varphi}} ^k$ obey $N(0,\sigma_{Z}^2)$\\
\qquad \textbf{end}\\
\qquad \% Create the overall population of parents and offspring: \\
\qquad \% ${{\cal P}_{ca}} = \left( {\mathop  \cup \limits_{i = 1}^{W} {{\cal P}_{{s_i}}}} \right) \cup {\cal P}$, 
$W = \left\{ {\begin{array}{*{20}{c}}
{W_0},\quad Z=0,\\
{W_{\max},\text{others}.}
\end{array}} \right.$\\
\qquad \% Select the best $W_{max}$ individuals as ${\textbf{p}}^{Z}$ for the DE procedure.\\
\qquad \% Note that ${\textbf{p}}^{Z}$ consists of $W_{max}$ weeds.\\
\qquad \%\% (\textbf{Operate the DE algorithm})\\
\qquad $\bm{\alpha} = \text{Mutation}({\textbf{p}}^{Z},\text{best},r_1,r_2,F_0)$, $\bm{\alpha} = \{\bm{\alpha}^i,i=1,\cdots,W_{\max} \}$;\\
\qquad ${\textbf{u}} = \text{Crossover}({\textbf{p}}^{Z},\bm{\alpha},C_r)$, ${\textbf{u}} = \{{\textbf{u}}^i,i=1,\cdots,W_{\max} \}$;\\
\qquad ${\textbf{p}}^Z = \text{Selection}({\textbf{u}}, {\textbf{p}}^{Z})$;\\
\qquad \% ${\textbf{p}}^Z$, taken as the new weed set ${\cal P}$  into the next generation. \\ 
\qquad $Z = Z + 1$;\\
\textbf{end}\\
\hline
\end{tabular*}
\end{table}
\par
To sum up, our QoS-guaranteed ALOHA-like random access algorithm relies on the IWO-DE algorithm to find the optimal access probability for each device.  
Each device transmits its \textit{traffic information} containing the traffic type (determined by the statistical property of the traffic) and the QoS requirement to the coordinator. The coordinator calculates devices' optimal access probabilities by solving the OP (\ref{eq-HQoS-OP-min}) with the help of the IWO-DE algorithm\footnote{The coordinator does not need to recalculate the devices' optimal access probabilities provided that the individual channel gains and \textit{traffic information} (i.e., the traffic type and QoS requirement) of these devices are unchanged. In this case, the IWO-DE is an off-line optimization algorithm for the problem considered. Hence, it does not make sense to incorporate the processing time of the IWO-DE algorithm into the statistical transmission delay considered, which is related with the underlying buffer size and the arrival information rate.}. 
Then, the coordinator broadcasts the optimum access probability vector for devices in the beacon frame.
Finally, each device \textit{distributedly} transmits its respective data to the coordinator with its unique optimum access probability.
\begin{table}[!t]
\renewcommand{\arraystretch}{1.2}
\caption{The Values of Parameters}
\centering
\label{table-Simu-para}
\centering
\begin{tabular}{l||c}
\hline
\bfseries Parameter & \bfseries Value\\
\hline
\hline
The number of devices: $N$ & 3-10\\
Available bandwidth for VLC system: $B$ & 20 [MHz]\\
Length of room & 10 [m]\\
Width of room & 20 [m]\\
Height of room & 5 [m]\\
Height of LED & 4.85 [m]\\
Transmitting power: $P_\text{t}$ & 100 [mW]\\
The detector responsivity: $\xi$ & 0.97 [A/W]\\
Semi-angle at half power: $\phi_{1/2}$ & 70 [deg.]\\
Width of the field of view: $\Psi_\text{C}$ & 70 [deg.]\\
Detector physical area of a PD: $A_0$ & 1.0 [$\text{cm}^2$]\\
Refractive index: $n_0$ & 1.5\\
Gain of an optical filter: $T_\text{s}$ & 0.53\\
Size of the initial population: $W_0$ & 60\\
Maximum no. of iterations: $Z_{\max}$ & 300\\
Maximum no. of weeds: ${W}_{\max}$ & 20/50/80\\
Maximum no. of offspring: ${S_{\max}}$ & 6\\
Minimum no. of offspring: ${S_{\min}}$ & 1\\
Non-linear modulation index: ${\phi}$ & 3\\
Initial standard deviation: ${\sigma_\text{initial}}$ & 0.15\\
Final standard deviation: ${\sigma_\text{final}}$ & $10^{-6}$\\
Scaling factor: $F_0$ & 0.75\\
Crossover probability: $C_\text{r}$ & 0.9\\
\hline
\end{tabular}
\end{table}
\section{Simulation Results and Discussions}
\label{sec-results}
In this paper, the coordinator uses the MMSE based SIC algorithm \cite{LinlinZhao2015} and the device having higher LOS channel gain is decoded first. 
We assume that the number of devices is $N=10$ except for the scenario of Fig. \ref{fig-IWODE-compare-throughput}.
Additionally, devices are uniformly distributed in the room. 
In the context of $M=2$, the distance between the PDs of the coordinator is $15$ cm. When $M=4$, the PDs constitute the $2 \times 2$ receiver array of $15 \; \text{cm} \times 15 \; \text{cm}$. 
We assume that the arrival process of the traffic of the device $j$ obeys the Poisson distribution with the parameter $\lambda_j$. 
$\lambda_j$ is also defined as the average arrival rate of the packet. 
The packet length is $1000$ bits, and the duration of the time slot is $0.5$ ms. 
The main system parameters and values used in our simulations are summarized in Table \ref{table-Simu-para}.

\subsection{Analytical versus Simulated EC of Individual Devices}
Our Monte Carlo simulation is based on $5 \times 10^5$ times repeated trials\footnote{An ALOHA-like random access VLC system was simulated, where the individual instantaneous service rates of $5 \times 10^5$ time slots were recorded for evaluating the EC according to (\ref{eq-EC}).}.
Table \ref{table-comparison-EC} shows the theoretical EC results and the Monte Carlo simulation results in the situation that all devices randomly pick their probabilities of being unblocked, access probabilities and QoS exponents.  
It is demonstrated that our analytical EC expression of (\ref{eq-ec-device-bits}) matches accurately with the results of the Monte Carlo simulation. 

\begin{figure}[tbp]
\centering
\includegraphics[width=3.5in]{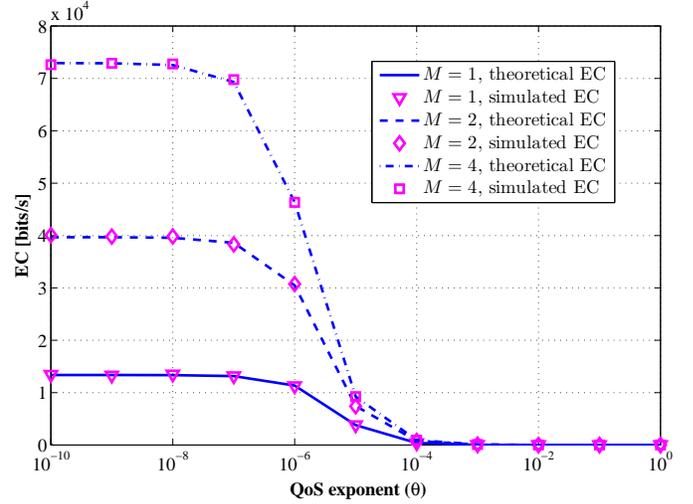} 
\caption{Comparison between theoretical ECs and the corresponding simulation results in the context of different MPR capabilities that are indicated by the values of $M$, assuming that the common probability of the LOS links being unblocked is $\beta=0.9$ and the access probability vector is $\textbf{p}=(1/N) \times {\bf{1}}_N^T$.}
\label{fig-exactEC-device-beta0.9}
\end{figure}

Fig. \ref{fig-exactEC-device-beta0.9} shows the EC comparison results when the coordinator has different MPR capabilities, indicated by $M$. 
It is demonstrated that the EC grows with the increase of the MPR capability $M$ of the coordinator.  
Fig. \ref{fig-exactEC-device-beta0.9} also indicates the accuracy of our analytical EC expression of (\ref{eq-ec-device-bits}). 
Additionally, it demonstrates that the EC decreases with the increase of the QoS exponent. 
The reason is that higher QoS exponent indicates more stringent delay QoS requirement. For a given service process, the maximum constant arrival rate that can be supported will decrease under stricter delay QoS requirement. 

\begin{table*}[!tbp]
\renewcommand{\arraystretch}{1.3}
\caption{Comparison between theoretical ECs and simulations in the random situation considered, $M=2$}
\label{table-comparison-EC}
\centering
\begin{tabular}{c||c|c|c|c|c|c|c|c|c|c}
\hline
\bfseries {Device ID} & \bfseries 1& \bfseries 2& \bfseries 3& \bfseries 4& \bfseries 5& \bfseries 6& \bfseries 7& \bfseries 8& \bfseries 9& \bfseries 10\\
\hline
\hline
\bfseries {Theoretical EC [bits/s]} & 5.27978 & 1362.97 & 0.295016 & 5306.56 & 9.87281 & 1225.59 & 610.200 & 26.8956 & 1167.75 & 1822.52\\
\bfseries {Simulated EC [bits/s]} & 5.27828 & 1272.98 & 0.294268 & 5346.41 & 9.77563 & 1225.77 & 572.928 & 26.7342 & 1169.21 & 1815.45\\
\hline
\end{tabular}
\end{table*}

\subsection{Performance of the QoS-Guaranteed ALOHA-like Random Access Algorithm}
\begin{figure}[tbp]
\centering
\includegraphics[width=3.5in]{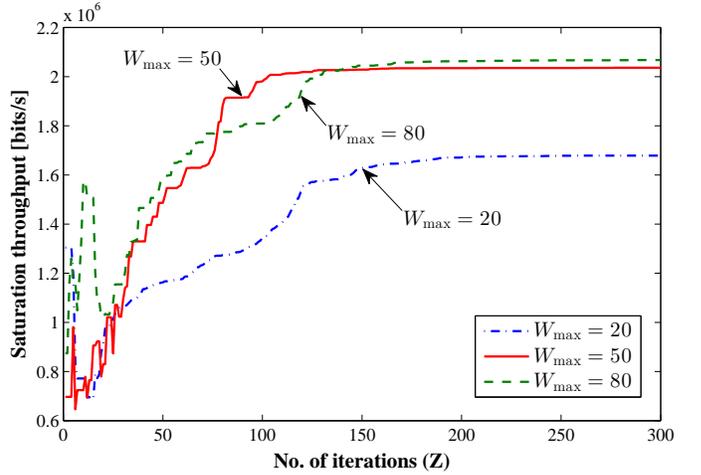} 
\caption{Performance of the IWO-DE algorithm with different values of the maximum weed size $W_\text{max}$ in the random situation considered.}
\label{fig-IWODE-Wmax-maxST}
\end{figure}

\begin{figure}[t]
\centering
\includegraphics[width=3.5in]{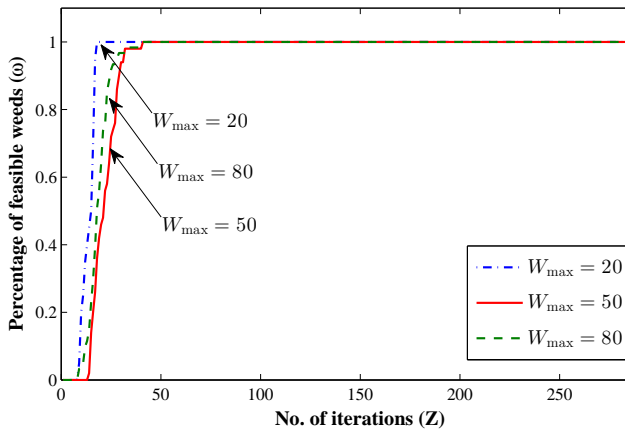} 
\caption{Percentage of feasible weeds for the IWO algorithm with different values of the maximum weed size $W_\text{max}$ in the random situation considered.}
\label{fig-IWODE-Wmax-omega}
\end{figure}

Fig. \ref{fig-IWODE-Wmax-maxST} and Fig. \ref{fig-IWODE-Wmax-omega} depict the maximum saturation throughput and the percentage of the feasible solutions of the IWO-DE algorithm with respect to different values of the maximum weed size $W_{\max}$ in the random situation, where the devices randomly pick their blocking status and the traffic parameters, such as the probability of being  unblocked ranging from $0.7$ to $1$, the arrival parameters ranging from $0.001$ to $0.01$, the QoS exponent ranging from $10^{-10}$ to $10^{-6}$. 
Obviously, increasing the maximum weed size from 20 to 80 improves the performance of the IWO-DE algorithm after convergence. 
Fig. \ref{fig-IWODE-Wmax-maxST} shows that the IWO-DE search algorithm is trapped into the local optimal solution when $W_{\max}=20$. 
This is because DE is a population-based method. The small maximum weed size  $W_{\max}$ limits the \textit{variation} of the DE procedure, and then limits the global search ability of the DE method. 
In Fig. \ref{fig-IWODE-Wmax-omega}, the curve of $W_{\max}=20$ is steeper than the others. The reason is that the population size of $W_{\max}=20$ is much lower than the other two situations. 
Fig. \ref{fig-IWODE-Wmax-maxST} and Fig. \ref{fig-IWODE-Wmax-omega} indicate that our IWO-DE algorithm has a desirable convergence when $W_{\max}=50$. 
Considering the convergence as well as the computational complexity, we choose $W_{\max}=50$ as the specific value used in the following simulations. 

\begin{figure}[t]
\centering
\includegraphics[width=3.5in]{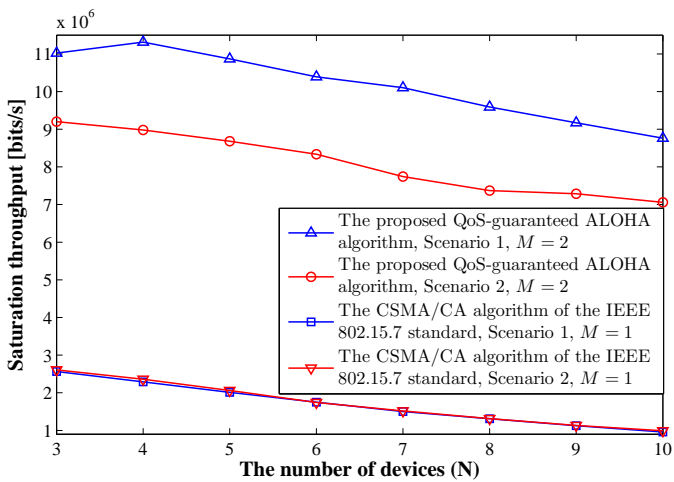} 
\caption{Comparison of the saturation throughput achieved by our QoS-guaranteed ALOHA-like random access algorithm and the CSMA/CA algorithm of the IEEE 802.15.7 standard. We assume that the common probability of the LOS links being unblocked is $\beta = 1$, the common QoS exponent is $\theta = 10^{-8}$, and the common arrival parameter is $\lambda = 0.01$.}
\label{fig-IWODE-compare-throughput}
\end{figure}

\par
Fig. \ref{fig-IWODE-compare-throughput} shows the comparison of the saturation throughput achieved by our QoS-guaranteed ALOHA-like random access algorithm and by the CSMA/CA algorithm of the IEEE 802.15.7 standard \cite{VLC-standard-15.7, VLC-survey-mac2015}. 
In Scenario 1 and Scenario 2, the difference in the channel gains of the devices is assumed to be large and small\footnote{To achieve this ``channel-gain difference'' goal in a fair manner for both algorithms, the actual locations of the devices are particularly selected from a larger set of uniformly distributed positions in both Scenario 1 and Scenario 2. More specifically, considering the benchmark single-packet reception, i.e., $M = 1$, we first calculate the individual instantaneous PHY transmission rates for all the elements of the larger position set, and then select two identical-size subsets of positions so that the channel-gain difference of one subset is large and the other is small, while making sure the average PHY transmission rate of both subsets is the same.  As a result, all the devices share a common access probability of $p = 1/N$ and for the CSMA/CA algorithm, the saturation throughput is equal in both scenarios, since it is insensitive to the channel-gain difference, as demonstrated by the results of Fig. \ref{fig-IWODE-compare-throughput}.}, respectively. 
In both scenarios, none of the devices is blocked, i.e., the channels are assumed to be ideal. Additionally, we assume that all the devices share the common QoS exponent constraint $\theta = 10^{-8}$ and the common arrival parameter $\lambda = 0.01$. Referring to \cite{VLC-Analysis-802.15.7}, we set the parameters of the CSMA/CA algorithm of the IEEE 802.15.7 standard as follows: $macMinBE = 3$, $macMaxBE=5$, $macMaxFrameRetries = 3$ and $macMaxCSMABackoffs = 4$. 
Furthermore, to make fair comparison, we also utilize Shannon's capacity formula to calculate the upper bound on the instantaneously achievable PHY transmission rate when quantifying the saturation throughput achieved by the CSMA/CA algorithm of the IEEE 802.15.7 standard.

\begin{figure}[t]
\centering
\includegraphics[width=3.5in]{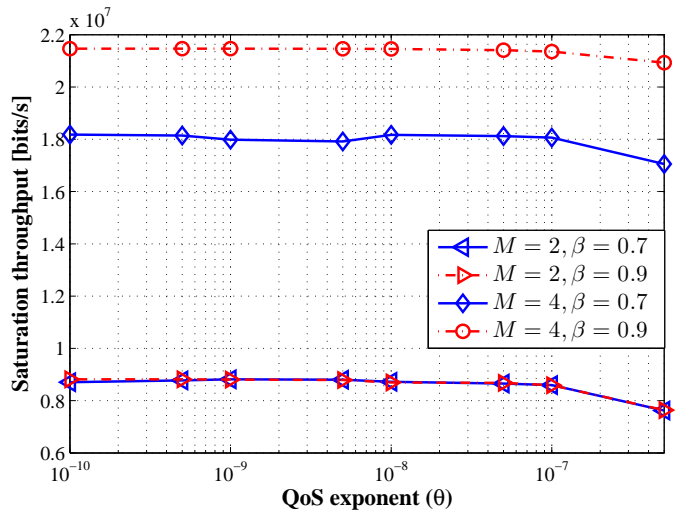} 
\caption{Performance of the proposed QoS-guaranteed ALOHA-like random access algorithm in different blocking situations, assuming the common arrival parameter as $\lambda = 0.01$.}
\label{fig-IWODE-beta-maxST}
\end{figure}

\par
As observed from Fig. \ref{fig-IWODE-compare-throughput}, the saturation throughput of our algorithm is far superior to that of the CSMA/CA algorithm, especially in Scenario 1, although the CSMA/CA algorithm benefits from the carrier sensing, which is assumed to be perfect in our simulations. This is because the MPR capability of the proposed algorithm reduces the chance of collisions in the random access procedure. More specifically, our algorithm adaptively adjusts the access probability vector of devices in order to fully utilize the MPR capability. Furthermore, there is no constraint on the minimum service rate of each device in our algorithm. As a result, the devices with higher channel gains have larger access probabilities in the context where they share a common delay QoS exponent value. In other words, the devices with higher channel gains have more chances to transmit data packets to the coordinator and thus a maximized saturation throughput of the VLC system is achieved.
Additionally, in our scheme the MPR capability of the coordinator is achieved by using the SIC algorithm, which performs better when there is a substantial difference in the channel gains of the simultaneously transmitting devices, provided that no power control is used \cite{Shaoshi2015}. Therefore, the gap of the saturation throughput between our algorithm and the benchmark CSMA/CA algorithm is much larger in Scenario 1 than in Scenario 2.
It is worth noting that our algorithm achieves the maximized saturation throughput of the VLC system by sacrificing the fairness, while the benchmark CSMA/CA algorithm guarantees that all devices have equal opportunities to transmit to the coordinator, thus achieving better fairness.
We also observe from Fig. \ref{fig-IWODE-compare-throughput} that the saturation throughput of both algorithms declines when the number of devices is increased. 
Nonetheless, the proposed algorithm remains substantially superior to the benchmark CSMA/CA algorithm in term of the saturation throughput achieved. 
Finally, our algorithm guarantees the statistical delay QoS requirements of devices, hence a marginal decrease in the saturation throughput with the increasing of the number of devices can be regarded as a reasonable cost for obtaining this benefit.
\par
In the scenario of Fig. \ref{fig-IWODE-beta-maxST}, all the devices share the common QoS constraint $\theta$ and the common probability of being unblocked $\beta$. 
Additionally, they share the common arrival parameter $\lambda=0.01$. 
Fig. \ref{fig-IWODE-beta-maxST} shows the maximum saturation throughput of our proposed ALOHA-like random access algorithm in different blocking situations. 
When the MPR capability is smaller, i.e., $M=2$, heavy collisions happens, the achievable maximum throughput of $\beta=0.7$ approaches that of $\beta=0.9$. 
Whereas the gap between the curve of $\beta=0.7$ and the curve of $\beta=0.9$ increases when $M=4$. 
This is because the stronger MPR capability alleviates collisions of the VLC system in a greater degree. In such a situation, the benefit of the less severely blocked channel condition becomes prominent. 
Fig. \ref{fig-IWODE-beta-maxST} also shows that the maximum saturation throughput only slightly decreases with the increase of the QoS exponent for the given $M$, which indicates that our QoS-guaranteed ALOHA-like random access algorithm is reliable when the QoS constraint varies. 
\begin{figure}[tbp]
\centering
\includegraphics[width=3.5in]{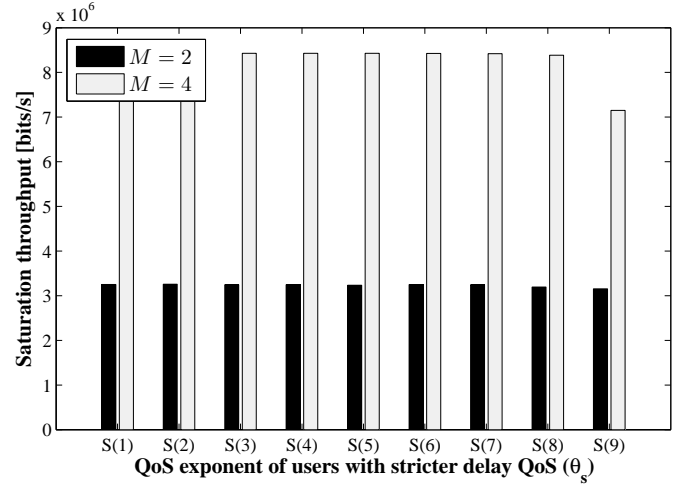} 
\caption{The maximum saturation throughput of our QoS-guaranteed ALOHA-like random access algorithm in  heterogeneous QoS situations, where ${\text{S}(2k-1)}$ ($k=1,\cdots,5$) represents the situation of $\theta_s=10^{k-11}$ and ${\text{S}(2k)}$ ($k=1,\cdots,4$) represents the situation of $\theta_s=5 \times 10^{k-11}$, while assuming that the common arrival parameter is $\lambda=0.01$, the common probability of devices being unblocked is $\beta=0.9$, and the common QoS exponent of the remaining half of the MPR-supported devices is $\theta_l=10^{-10}$.}
\label{fig-IWODE-QoS2-maxST}
\end{figure}

\begin{figure}[tbp]
\centering
\includegraphics[width=3.5in]{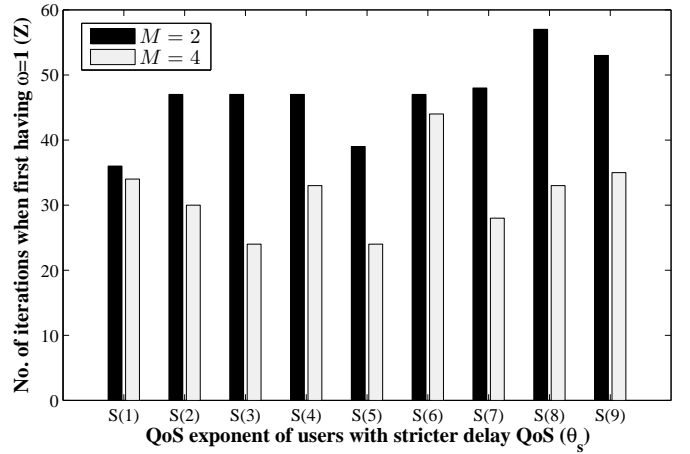} 
\caption{ The number of iterations when we first have $\omega=1$ in heterogeneous QoS situations, where ${\text{S}(2k-1)}$ ($k=1,\cdots,5$) represents the situation of $\theta_s=10^{k-11}$ and ${\text{S}(2k)}$ ($k=1,\cdots,4$) represents the situation of $\theta_s=5 \times 10^{k-11}$. We also assume that $\lambda=0.01$, $\beta=0.9$, $\theta_l=10^{-10}$, which are the same as those of Fig. \ref{fig-IWODE-QoS2-maxST}.}
\label{fig-IWODE-QoS2-feasible0}
\end{figure}

\par
In the context of Fig. \ref{fig-IWODE-QoS2-maxST} and Fig. \ref{fig-IWODE-QoS2-feasible0}, the devices are divided into two groups, where $\displaystyle \frac{M}{2}$ devices with the stricter QoS constraint $\theta_s$ are in one of the groups, while the remaining devices with $\theta_l = 10^{-10}$ in the other group. 
${\text{S}(2k-1)}$ ($k=1,\cdots,5$) represents the situation of $\theta_s=10^{k-11}$ and ${\text{S}(2k)}$ ($k=1,\cdots,4$) represents the situation of $\theta_s=5 \times 10^{k-11}$. 
Additionally, they share the common arrival parameter $\lambda=0.01$. 
Fig. \ref{fig-IWODE-QoS2-maxST} shows the maximum saturation throughput of our QoS-guaranteed ALOHA-like random access algorithm versus $\theta_s$. 
It is demonstrated that the performance of our QoS-guaranteed ALOHA-like random access algorithm remains favourable even when the QoS constraints become stricter, i.e. when $\theta_s$ becomes larger.  
Additionally, the maximum saturation throughput is doubled when the MPR capability $M$ increases from $2$ to $4$, which indicates that our algorithm fully utilizes the MPR capability to improve the performance of the system  considered. 
Fig. \ref{fig-IWODE-QoS2-feasible0} shows the value of the generation number $Z$ versus $\theta_s$ when all candidate solutions of the IWO-DE algorithm start to be feasible. 
The fluctuations in Fig. \ref{fig-IWODE-QoS2-feasible0} are caused by the randomness inherent in the IWO-DE algorithm, such as the randomly generated  offspring in the \textit{spatial dispersion} step and the random  \textit{variations} in the DE procedure. 
Obviously, it is harder to search for the feasible access probability vector when $M=2$, which matches the fact that it is harder to guarantee the QoS when the MPR capability $M$ is lower.

\begin{figure}[tbp]
\centering
\includegraphics[width=3.5in]{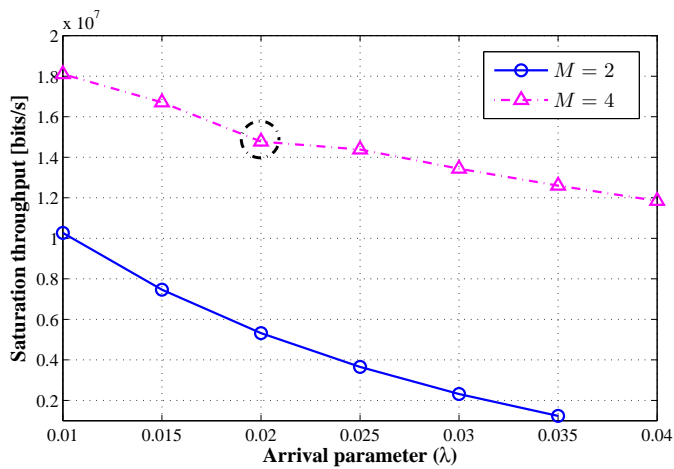} 
\caption{The maximum saturation throughput of our QoS-guaranteed ALOHA-like random access algorithm versus the arrival parameters $\lambda$. We assume $\beta=0.9$, $\theta_l=10^{-10}$ and $\theta_s=10^{-7}$, as defined in a similar fashion to those of Fig. \ref{fig-IWODE-QoS2-maxST} and Fig. \ref{fig-IWODE-QoS2-feasible0}.}
\label{fig-IWODE-lambda-maxST}
\end{figure}

\par
In the scenario of Fig. \ref{fig-IWODE-lambda-maxST}, all devices share the common probability of being unblocked, i.e., $\beta=0.9$. Again, the devices are divided into two groups, where $\displaystyle \frac{M}{2}$ of them with the stricter QoS constraint $\theta_s=10^{-7}$ are in one of the groups, while the remaining devices with looser QoS constraint $\theta_l=10^{-10}$ in the other group. 
Fig. \ref{fig-IWODE-lambda-maxST} shows the maximum saturation throughput of our proposed ALOHA-like random access algorithm versus the common arrival parameter $\lambda$. 
The maximum saturation throughput declines with the increase of $\lambda$ in the situation that $M=2$, so does in the situation that $M=4$.  
But the curve of $M=4$ falls more slowly, especially after the noted point. 
The increase of $\lambda$ makes it more difficult to satisfy the first constraint of the OP (\ref{eq-HQoS-OP-min}).  
Thus the set of the feasible solutions of (\ref{eq-HQoS-OP-min}) becomes smaller. As a result, the achievable maximum saturation throughput becomes lower. 
Furthermore, there is no feasible solution to the OP (\ref{eq-HQoS-OP-min}) when $M=2$ and $\lambda = 0.04$. In other words, the QoS requirement cannot be guaranteed in this situation. 
Furthermore, after the noted point in Fig. \ref{fig-IWODE-lambda-maxST} the maximum saturation throughput of $M=4$ is more than twice that of $M=2$. 
It is demonstrated that increasing the MPR capability of the coordinator obtains remarkable performance advantage when $\lambda$ is larger. 

\section{Conclusion}
\label{sec-conclusion}
In this paper, upon considering the influence of the MIMO-SIC algorithm in the PHY layer, we have explored the ALOHA-like distributed random access algorithm for the MPR-aided uplink VLC system with heterogeneous QoS guarantees. 
We modelled the system as a conflict graph and proposed the concept of \textit{feasible access state} to characterise the random group of concurrently transmitting devices, which is resulted from the random access.   
Relying on the probability of the \textit{feasible access state} and the feature of the MIMO-SIC algorithm, we analysed the instantaneously achievable transmission rate and its probability distribution for each device. Based on the EC theory, we  derived the EC of each device to model its statistical delay QoS constraint. 
Then, we formulated the random access problem as a saturation throughput maximization problem subject to multiple statistical delay QoS constraints. We  converted the resultant non-concave COP to the unconstrained bi-objective OP and obtained the optimal access probability vector with the help of the IWO-DE algorithm.
The impact of a wide range of relevant system parameters have been investigated based on our analysis and simulation results. 

%
\appendices
\section{Proof of Theorem 3.1}
\label{app-theorem-non-convex}
Let us employ the Hessian matrix for examining the non-concavity of the function $\eta(\textbf{p})$, which is given by:
\begin{equation}
\label{eq-Hessian}
\cal{H}(\eta({\textbf{p}})) = \left[ {\begin{array}{*{20}{c}}
{\frac{{{\partial ^2}\eta(\textbf{p}) }}{{\partial p_1^2}}}&{\frac{{{\partial ^2}\eta(\textbf{p}) }}{{\partial {p_1}\partial {p_2}}}}& \cdots &{\frac{{{\partial ^2}\eta(\textbf{p}) }}{{\partial {p_1}\partial {p_N}}}}\\
{\frac{{{\partial ^2}\eta(\textbf{p}) }}{{\partial {p_2}\partial {p_2}}}}&{\frac{{{\partial ^2}\eta(\textbf{p}) }}{{\partial p_2^2}}}& \cdots &{\frac{{{\partial ^2}\eta(\textbf{p}) }}{{\partial {p_2}\partial {p_N}}}}\\
 \vdots & \vdots & \ddots & \vdots \\
{\frac{{{\partial ^2}\eta(\textbf{p}) }}{{\partial {p_N}\partial {p_1}}}}&{\frac{{{\partial ^2}\eta(\textbf{p}) }}{{\partial {p_N}\partial {p_2}}}}& \cdots &{\frac{{{\partial ^2}\eta(\textbf{p}) }}{{\partial p_N^2}}}
\end{array}} \right].
\end{equation}
\begin{equation}
\begin{array}{l}
\displaystyle
\frac{\partial \eta(\textbf{p})}{\partial p_n}=\sum\limits_{\bm{\sigma}^i \in {\cal I}} \sum\limits_{j=1}^{N}  R_j^i \sigma_j^i \pi_{(-n)}^i [(1-\sigma_n^i)(-\beta_n)+\sigma_n^i \beta_n]\\
\quad\quad\quad = \sum\limits_{\bm{\sigma}^i \in {\cal I}} (2\sigma_n^i-1)\beta_n \sum\limits_{j=1}^{N}  R_j^i \sigma_j^i \pi_{(-n)}^i,
\end{array}
\end{equation}
where $\pi_{(-n)}^i=\prod\limits_{k=1,k \neq n}^N {\left[{(1-\sigma_k^i)(1-p_k\beta_k)+\sigma_k^ip_k\beta_k}\right]}$,  
and $\displaystyle \frac{{\partial ^2}\eta(\textbf{p})}{\partial p_n^2} = 0, n = 1,2,\cdots,N $. We know that the sum of the eigenvalues equals the sum of the elements on the main diagonal. Furthermore, it is obvious that the function $\eta(\textbf{p})$ is not linear with respect to the access probability vector ${\textbf{p}}$. Hence, the eigenvalues of the Hessian matrix are not all non-positive. Thus $\eta(\textbf{p})$ is not concave with respect to $\textbf{p}$. So is the EC function $\mathsf{EC}_j(\textbf{p};\theta_j)$. As a result, the OP (\ref{eq-HQoS-OP}) is neither concave nor convex. 



\ifCLASSOPTIONcaptionsoff
  \newpage
\fi



\bibliographystyle{IEEEtran}
\bibliography{IEEEabrv,mpr-references}

\begin{thebibliography}{10}
\providecommand{\url}[1]{#1}
\csname url@samestyle\endcsname
\providecommand{\newblock}{\relax}
\providecommand{\bibinfo}[2]{#2}
\providecommand{\BIBentrySTDinterwordspacing}{\spaceskip=0pt\relax}
\providecommand{\BIBentryALTinterwordstretchfactor}{4}
\providecommand{\BIBentryALTinterwordspacing}{\spaceskip=\fontdimen2\font plus
\BIBentryALTinterwordstretchfactor\fontdimen3\font minus
  \fontdimen4\font\relax}
\providecommand{\BIBforeignlanguage}[2]{{%
\expandafter\ifx\csname l@#1\endcsname\relax
\typeout{** WARNING: IEEEtran.bst: No hyphenation pattern has been}%
\typeout{** loaded for the language `#1'. Using the pattern for}%
\typeout{** the default language instead.}%
\else
\language=\csname l@#1\endcsname
\fi
#2}}
\providecommand{\BIBdecl}{\relax}
\BIBdecl

\bibitem{VLC-5G-2014}
S.~Wu, H.~Wang, and C.~Youn, ``Visible light communications for {5G} wireless
  networking systems: From fixed to mobile communications,'' \emph{{IEEE}
  Netw.}, vol.~{28}, no.~{6}, pp. {41--45}, Nov./Dec. {2014}.

\bibitem{VLC-home}
M.~Chen, J.~Wan, S.~Gonzalez, X.~Liao, and V.~Leung, ``A survey of recent
  developments in home {M2M} networks,'' \emph{{IEEE} Commun. Surveys Tuts.},
  vol.~{16}, no.~{1}, pp. {98--114}, 1st Quart. {2014}.

\bibitem{5G-M2M-survey}
M.~Agiwal, A.~Roy, and N.~Saxena, ``Next generation {5G} wireless networks: A
  comprehensive survey,'' \emph{{IEEE} Commun. Surveys Tuts.}, vol.~PP, no.~99,
  pp. 1--1, 2016.

\bibitem{VLC-uplink-TDD}
Y.~Liu, C.~Yeh, C.~Chow, Y.~Liu, Y.~Liu, and H.~Tsang, ``Demonstration of
  bi-directional {LED} visible light communication using {TDD} traffic with
  mitigation of reflection interference,'' \emph{Opt. Express}, vol.~{20},
  no.~{21}, pp. {23\,019--23\,024}, Oct. {2012}.

\bibitem{VLC-uplink-Arch}
Y.~Wang, N.~Chi, Y.~Wang, L.~Tao, and J.~Shi, ``Network architecture of a
  high-speed visible light communication local area network,'' \emph{{IEEE}
  Photon. Technol. Lett.}, vol.~27, no.~2, pp. 197--200, Jan. 2015.

\bibitem{VLC-standard-15.7}
\emph{{IEEE} Standard for Local and Metropolitan Area Networks--Part 15.7:
  Short-Range Wireless Optical Communication Using Visible Light}, IEEE Std.
  802.15.7-2011, Sep. 2011.

\bibitem{VLC-Analysis-802.15.7}
S.~Nobar, K.~Mehr, and J.~Niya, ``Comprehensive performance analysis of {IEEE}
  802.15.7 {CSMA/CA} mechanism for saturated traffic,'' \emph{{J. Opt. Commun.
  Netw.}}, vol.~{7}, no.~{2}, pp. {62--73}, Feb. {2015}.

\bibitem{VLC-uplink-csmacd}
Q.~Wang and D.~Giustiniano, ``Intra-frame bidirectional transmission in
  networks of visible {LED}s,'' \emph{{IEEE/ACM} Trans. Netw.}, 2016 (early
  access in IEEE Xplore).

\bibitem{VLC-M2M}
O.~Ergul, E.~Dinc, and O.~Akan, ``Communicate to illuminate: State-of-the-art
  and research challenges for visible light communications,'' \emph{Phys.
  Commun.}, vol.~17, pp. 72--85, Dec. 2015.

\bibitem{VLC-LOS-channel}
T.~Komine and M.~Nakagawa, ``Fundamental analysis for visible-light
  communication system using {LED} lights,'' \emph{{IEEE} Trans. Consum.
  Electron.}, vol.~{50}, no.~{1}, pp. {100--107}, Feb. {2004}.

\bibitem{VLC-LAST2015}
M.~Biagi, S.~Pergoloni, and A.~Vegni, ``{LAST}: A framework to localize,
  access, schedule, and transmit in indoor {VLC} systems,'' \emph{J. Lightw.
  Technol.}, vol.~{33}, no.~{9}, pp. {1872--1887}, May {2015}.

\bibitem{MPR-LangTong2013}
D.~Chan, T.~Berger, and L.~Tong, ``Carrier sense multiple access communications
  on multipacket reception channels: Theory and applications to {IEEE} 802.11
  wireless networks,'' \emph{{IEEE} Trans. Commun.}, vol.~61, no.~1, pp.
  266--278, Jan. 2013.

\bibitem{Shaoshi2015}
S.~Yang and L.~Hanzo, ``Fifty years of {MIMO} detection: The road to
  large-scale {MIMO}s,'' \emph{{IEEE} Commun. Surveys Tuts.}, vol.~17, no.~4,
  pp. 1941--1988, 4th Quart. 2015.

\bibitem{LinlinZhao2015}
L.~Zhao, X.~Chi, P.~Li, and L.~Guan, ``A {MPR} optimization algorithm for {FSO}
  communication system with star topology,'' \emph{Opt. Commun.}, vol. 356, pp.
  147--154, Dec. 2015.

\bibitem{MPR-zigzag}
S.~Gollakota and D.~Katabi, ``{ZigZag} decoding: Combating hidden terminals in
  wireless networks,'' \emph{{ACM} SIGCOMM Computer Communication Review},
  vol.~38, no.~4, pp. 159--170, Aug. 2008.

\bibitem{MPR-MAC-backoff2009}
Y.~Zhang, P.~Zheng, and S.~Liew, ``How does multiple-packet reception
  capability scale the performance of wireless local area networks?''
  \emph{{IEEE} Trans. Mobile Comput.}, vol.~8, no.~7, pp. 923--935, Jul. 2009.

\bibitem{MPR-multi-round}
Y.~Zhang, ``Multi-round contention in wireless {LAN}s with multipacket
  reception,'' \emph{{IEEE} Trans. Wireless Commun.}, vol.~9, no.~4, pp.
  1503--1513, Apr. 2010.

\bibitem{MPR-Access-YunH2014}
Y.~Bae, B.~Choi, and A.~Alfa, ``Achieving maximum throughput in random access
  protocols with multipacket reception,'' \emph{{IEEE} Trans. Mobile Comput.},
  vol.~13, no.~3, pp. 497--511, Mar. 2014.

\bibitem{MIMO-CSMA2014}
S.~Wu, W.~Mao, and X.~Wang, ``Performance study on a {CSMA/CA}-based {MAC}
  protocol for multi-user {MIMO} wireless {LAN}s,'' \emph{{IEEE} Trans.
  Wireless Commun.}, vol.~13, no.~6, pp. 3153--3166, Jun. 2014.

\bibitem{MPR-fengfeng}
H.~Yu, X.~Chi, and J.~Liu, ``An integrated {PHY}-{MAC} analytical model for
  {IEEE} 802.15.7 {VLC} network with {MPR} capability,'' \emph{Opt. Lett.},
  vol.~10, no.~5, pp. 365--368, Sep. 2014.

\bibitem{EA1992}
C.~Chang and J.~Thomas, ``Effective bandwidth in high-speed digital networks,''
  \emph{{IEEE} J. Sel. Areas Commun.}, vol.~13, no.~6, pp. 1091--1100, Aug.
  1995.

\bibitem{EC2003}
D.~Wu and R.~Negi, ``Effective capacity: A wireless link model for support of
  quality of service,'' \emph{{IEEE} Trans. Wireless Commun.}, vol.~2, no.~4,
  pp. 630--644, Jul. 2003.

\bibitem{IWO-EC}
A.~Aijaz, M.~Tshangini, M.~Nakhai, X.~Chu, and A.~Aghvami, ``Energy-efficient
  uplink resource allocation in {LTE} networks with {M2M}/{H2H} co-existence
  under statistical {QoS} guarantees,'' \emph{{IEEE} Trans. Commun.},
  vol.~{62}, no.~{7}, pp. {2353--2365}, Jul. {2014}.

\bibitem{EC-LTE2015}
Y.~Li, L.~Liu, H.~Li, J.~Zhang, and Y.~Yi, ``Resource allocation for
  delay-sensitive traffic over {LTE}-advanced relay networks,'' \emph{{IEEE}
  Trans. Wireless Commun.}, vol.~14, no.~8, pp. 4291--4303, Aug. 2015.

\bibitem{Hanzo2015}
F.~Jin, R.~Zhang, and L.~Hanzo, ``Resource allocation under delay-guarantee
  constraints for heterogeneous visible-light and {RF} femtocell,''
  \emph{{IEEE} Trans. Wireless Commun.}, vol.~14, no.~2, pp. 1020--1034, Feb.
  2015.

\bibitem{EC-performance2015}
S.~Efazati and P.~Azmi, ``Statistical quality of service provisioning in
  multi-user centralised networks,'' \emph{IET. Commun.}, vol.~9, no.~5, pp.
  621--629, Mar. 2015.

\bibitem{Smith1980Receiver}
R.~Smith and S.~Personick, ``Receiver design for optical fiber communication
  systems,'' in \emph{Semiconductor Devices for Optical Communication},
  H.~Kressel, Ed.\hskip 1em plus 0.5em minus 0.4em\relax New York:
  Springer-Verlag, 1980.

\bibitem{FWC2005}
D.~Tse and P.~Viswanath, \emph{Fundamentals of Wireless Communication}.\hskip
  1em plus 0.5em minus 0.4em\relax New York, USA: Cambridge University Press,
  2005.

\bibitem{IWO-2006}
A.~Mehrabian and C.~Lucas, ``A novel numerical optimization algorithm inspired
  from weed colonization,'' \emph{{Ecol. Inform.}}, vol.~{1}, no.~{4}, pp.
  {355--366}, Dec. {2006}.

\bibitem{IWO-DE2013}
X.~Cai, Z.~Hu, and Z.~Fan, ``A novel memetic algorithm based on invasive weed
  optimization and differential evolution for constrained optimization,''
  \emph{{Soft Computing}}, vol.~{17}, no.~{10}, pp. {1893--1910}, Oct. {2013}.

\bibitem{book-convex}
S.~Boyd and L.~Vandenberghe, \emph{Convex Optimization}.\hskip 1em plus 0.5em
  minus 0.4em\relax New York, USA: Cambridge University Press, 2004.

\bibitem{VLC-survey-mac2015}
P.~Pathak, X.~Feng, P.~Hu, and P.~Mohapatra, ``Visible light communication,
  networking, and sensing: A survey, potential and challenges,'' \emph{{IEEE}
  Commun. Surveys Tuts.}, vol.~17, no.~4, pp. 2047--2077, 4th Quart. 2015.

\end{thebibliography}
%



%




\begin{IEEEbiography}[{\includegraphics[width=1in,height=1.25in,clip,keepaspectratio]{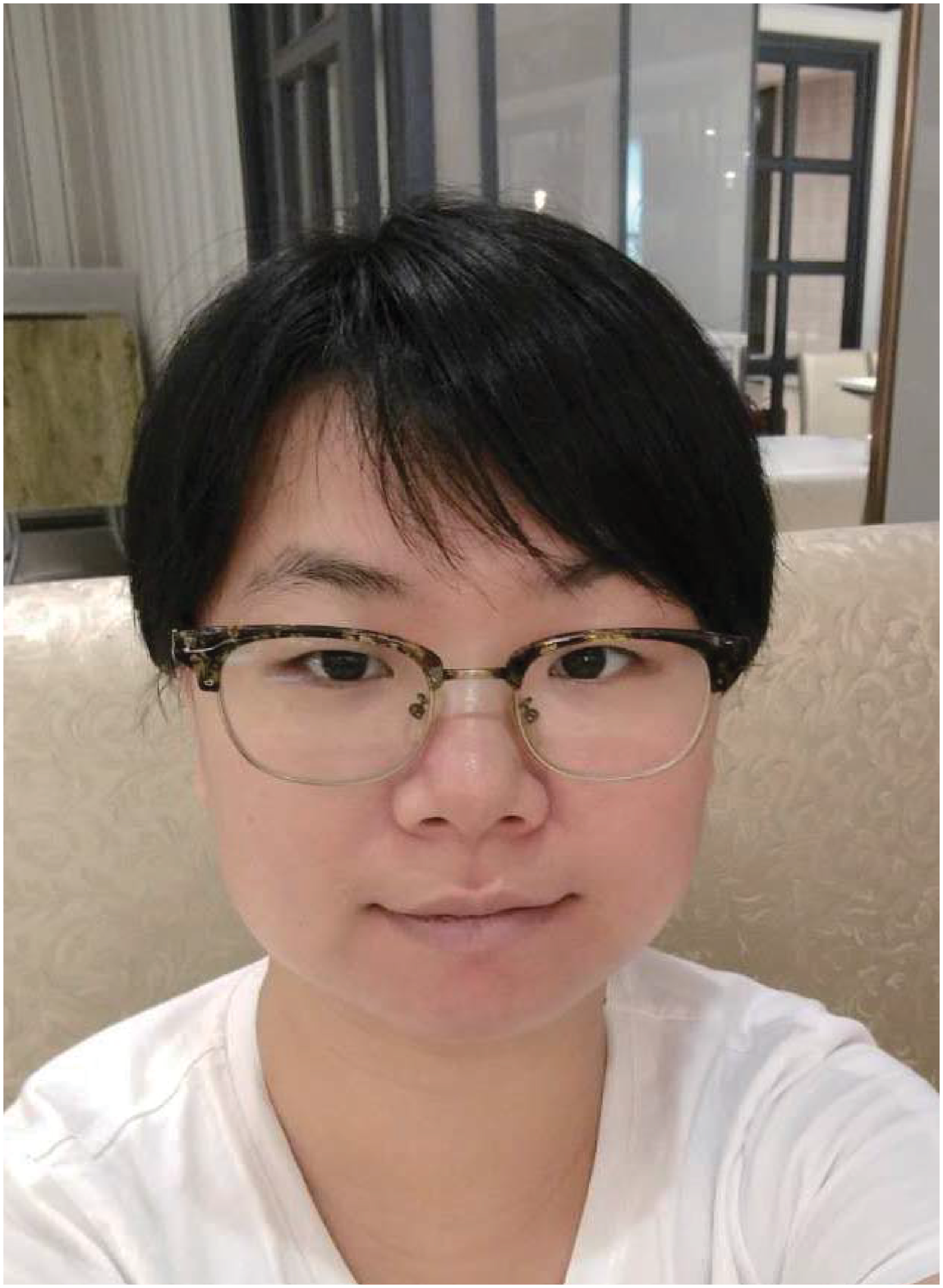}}]{Linlin Zhao}
received her B.Eng. and M.S. degrees from the Department of Communication Engineering, Jilin University, Changchun, China in 2009 and 2012, respectively. She is currently pursuing the Ph.D. degree in the Department of Communication Engineering, Jilin University, Changchun, China. Her research interests include the throughput optimal random access algorithms, resource allocation schemes, delay analysis and optimization in random access aided wireless networks.
\end{IEEEbiography}
\begin{IEEEbiography}[{\includegraphics[width=1in,height=1.25in,clip,keepaspectratio]{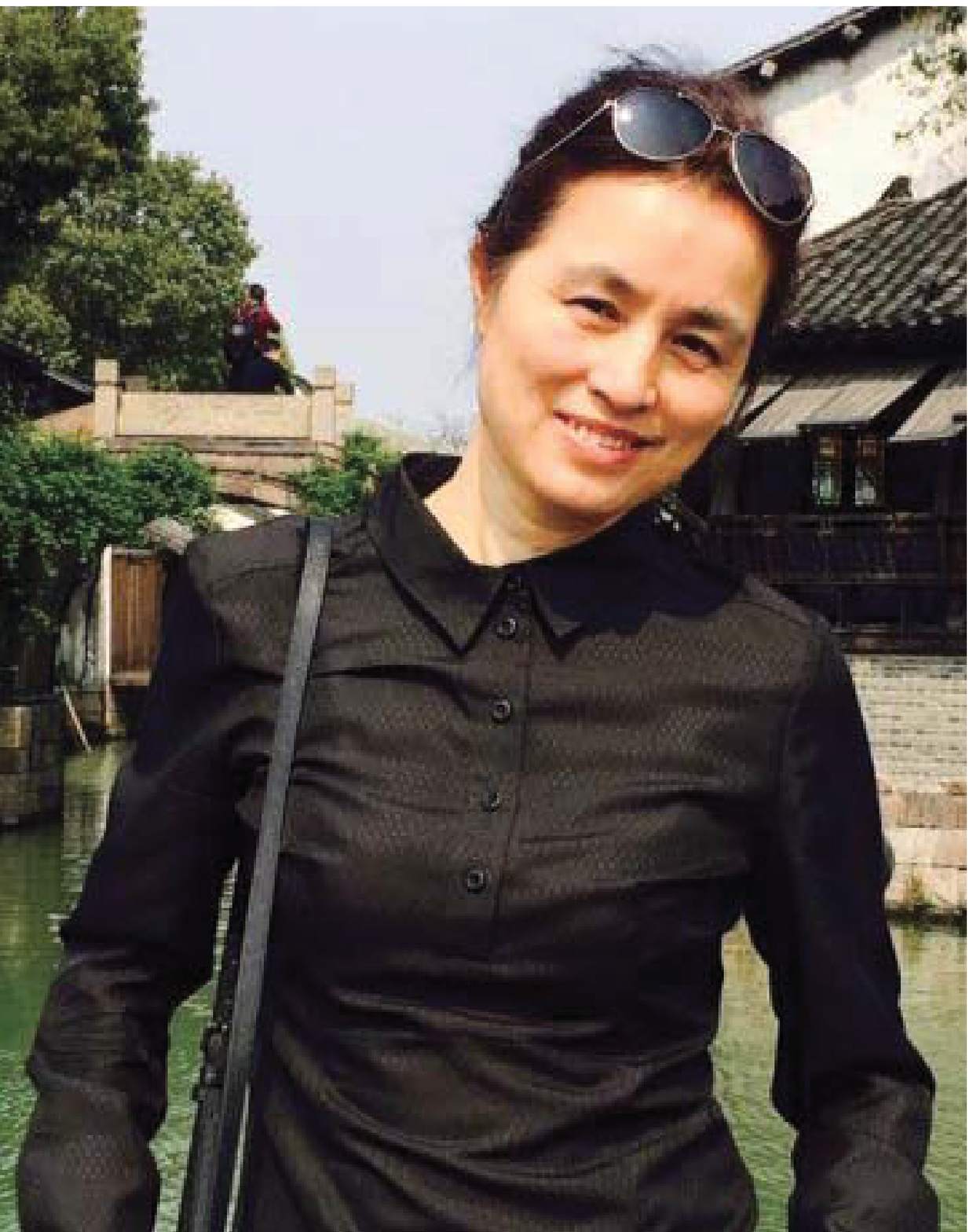}}]{Xuefen Chi}
received her B.Eng. degree in Information Engineering from Beijing University of Posts and Telecommunications (BUPT), Beijing, China in Jul. 1984. She received the M.S. and Ph.D. degree from Changchun Institute of Optics, Fine Mechanics and Physics, Chinese Academy of Sciences, Changchun, China, in 1990 and 2003 respectively. She was a visiting scholar at Department of Computer Science, Loughborough University, England, UK, in 2007, and School of Electronics and Computer Science, University of Southampton, Southampton, UK, in 2015. Currently, she is a professor at the Department of Communication Engineering, Jilin University, China. Her research interests include machine type communication, indoor visible light communication, random access algorithms, delay-QoS guarantees, queuing theory and its applications.
\end{IEEEbiography}
\begin{IEEEbiography}[{\includegraphics[width=1in,height=1.25in,clip,keepaspectratio]{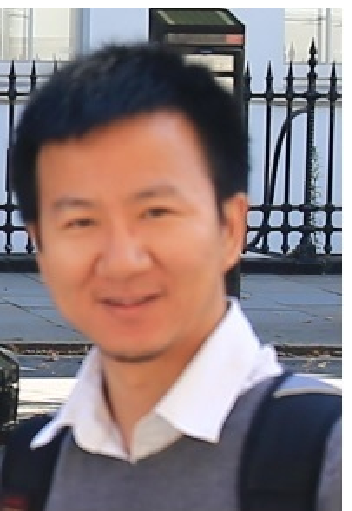}}]{Shaoshi Yang}
received his B.Eng. degree in Information Engineering from Beijing University of Posts and Telecommunications (BUPT),
Beijing, China in Jul. 2006, his first Ph.D. degree in Electronics and Electrical Engineering from University of Southampton, U.K. in Dec. 2013, and his second Ph.D. degree in Signal and Information Processing from BUPT in Mar. 2014. He is now working as a Postdoctoral Research Fellow in University of Southampton, U.K. From November 2008 to February 2009, he was an Intern Research Fellow with the Communications Technology Lab (CTL), Intel Labs, Beijing,
China, where he focused on Channel Quality Indicator Channel (CQICH) design for mobile WiMAX (802.16m) standard. His research interests include MIMO signal processing, green radio, heterogeneous networks, cross-layer
interference management, convex optimization and its applications. He has published in excess of 30 research papers on IEEE journals. Shaoshi has received a number of academic and research awards, including the prestigious Dean’s Award for Early Career Research Excellence at University of Southampton, the PMC-Sierra Telecommunications Technology Paper Award at BUPT, the Electronics and Computer Science (ECS) Scholarship of University of Southampton, and the Best PhD Thesis Award of BUPT. He is a member of IEEE/IET, and a junior member of Isaac Newton Institute for Mathematical Sciences, Cambridge University, U.K. He also serves as a TPC member of several major IEEE conferences, including IEEE ICC, GLOBECOM, VTC, WCNC, PIMRC, ICCVE, HPCC, and as a Guest Associate Editor of IEEE Journal on Selected Areas in Communications.
(http://sites.google.com/site/shaoshiyang/)
\end{IEEEbiography}




\end{document}